\documentclass[11pt]{article}
\usepackage{jcapmod}
\usepackage{rotating}
\usepackage{amsmath}
\usepackage{mathtools}
\usepackage{amsthm}
\usepackage{amsfonts}
\usepackage{graphicx}
\usepackage{psfrag}
\usepackage{hyperref}
\usepackage{amssymb}
\usepackage{lscape}
\usepackage{slashed}
\usepackage{lscape}
\usepackage{color}
\usepackage{url}
\usepackage{caption}
\usepackage{bm}
\usepackage{subfig}
\usepackage{epsfig}

\newcommand{\mg}[1]{\textcolor{blue}{#1}}

\newcommand{\beq}{\begin{equation}}
\newcommand{\eeq}{\end{equation}}
\newcommand{\bea}{\begin{eqnarray}}
\newcommand{\eea}{\end{eqnarray}}
\newcommand{\ga}{\lower.7ex\hbox{$\;\stackrel{\textstyle>}{\sim}\;$}}
\newcommand{\la}{\lower.7ex\hbox{$\;\stackrel{\textstyle<}{\sim}\;$}}

\newcommand{\eff}{{\rm{\it eff}}}
\newcommand{\bp}{\boldsymbol{p}}
\newcommand{\bk}{\boldsymbol{k}}

\makeatletter
\newcommand{\Cen}[2]{%
  \ifmeasuring@
    #2%
  \else
    \makebox[\ifcase\expandafter #1\maxcolumn@widths\fi]{$\displaystyle#2$}%
  \fi
}
\makeatother

\begin{document}

\begin{flushright}
UMN--TH--4006/20, FTPI--MINN--20/37 \\
IFT-UAM/CSIC-20-185 \\
KIAS-P20071
\end{flushright}
%%%%%%%%%%%%%%%%%%%%%% F I G U R E %%%%%%%%%%%%%%%%%%%%%%%%%%%%%%%%%%%

\vspace{0.5cm}
\begin{center}
%{\bf {\large Radiative inflaton and dark matter}}
{\bf {\large Inflaton Oscillations and Post-Inflationary Reheating}}
\end{center}

\vspace{0.05in}

\begin{center}{%\large
{\bf Marcos~A.~G.~Garcia}$^{a,b}$,
{\bf Kunio Kaneta}$^{c}$,
{\bf Yann Mambrini}$^{d}$}, and
{\bf Keith~A.~Olive}$^{e,f}$
\end{center}

\begin{center}
 {\em $^a$Instituto de F\'isica Te\'orica (IFT) UAM-CSIC, Campus de Cantoblanco, 28049, Madrid, Spain}\\[0.2cm] 
 {\em $^b$Departamento de F\'isica Te\'orica, Universidad Aut\'onoma de Madrid (UAM), Campus de Cantoblanco, 28049 Madrid, Spain}\\[0.2cm] 
 {\em $^c$School of Physics, Korea Institute for Advanced Study, Seoul 02455, Korea}\\[0.2cm]
  {\em $^d$ Universit\'e Paris-Saclay, CNRS/IN2P3, IJCLab, 91405 Orsay, France}\\[0.2cm] 
 {\em $^e$William I. Fine Theoretical Physics Institute, School of
 Physics and Astronomy, University of Minnesota, Minneapolis, MN 55455,
 USA}\\[0.2cm] 
{\em $^f$School of
 Physics and Astronomy, University of Minnesota, Minneapolis, MN 55455,
 USA}
 
\end{center}

\bigskip

\centerline{\bf ABSTRACT}

\noindent  
We analyze in detail the perturbative decay of the inflaton oscillating about a generic form of its potential $V(\phi) = \phi^k$, taking into account the effects of non-instantaneous reheating. We show that evolution of the temperature as a function of the cosmological scale factor depends on the spin statistics of the final state decay products when $k > 2$. 
We also include the inflaton-induced mass of the final states  leading to either kinematic suppression or enhancement if the final states are fermionic or bosonic respectively.  We compute the maximum temperature reached after inflation, the subsequent evolution of the temperature and the final reheat temperature.  We apply our results to the computation of the dark matter abundance through thermal scattering during reheating. We also provide an
example based on supersymmetry for the coupling of the inflaton to matter. 

\vspace{0.2in}

\begin{flushleft}
December 2020
\end{flushleft}
\medskip
\noindent

\newpage

\hypersetup{pageanchor=true}

\tableofcontents

\newpage

%%%%%%%%%%%%%%%%%%%%%%%%%%%
\section{Introduction}
%%%%%%%%%%%%%%%%%%%%%%%%%%%

The inflationary paradigm \cite{reviews} is well ensconced in the standard model of modern cosmology. Specific models of inflation can be tested by observations, 
most notably by the anisotropy spectrum of the cosmic microwave background (CMB) \cite{planckinf}.
A necessary feature of all inflationary models is the ability to amply reheat the universe
following the period of exponential expansion, leading to a radiation dominated epoch.
Often, perturbative reheating occurs as the inflaton begins a series of oscillations
about a minimum. When massive inflaton oscillations dominate the energy density,
the universe expands as if it were matter dominated until the inflaton decays
to relativistic particles which thermalize and reheat the Universe \cite{dl,nos}. 

A commonly used approximation to reheating is a pair of assumptions:
instantaneous decay and instantaneous thermalization. 
There has been a substantial amount of work which takes into account non-instantaneous reheating \cite{Giudice:2000ex,Garcia:2017tuj,grav2,Chen:2017kvz,Garcia:2020eof,Bernal:2020gzm,Co:2020xaf} or thermalization \cite{Davidson:2000er,Harigaya:2013vwa,Mukaida:2015ria,GA} after inflation. In this work, we maintain the instantaneous thermalization approximation, but consider in detail the evolution of the reheat process for general decays of the inflaton.
It is common to assume that after inflation,
a massive inflaton begins oscillating about a minimum.
As decays begin, the decay products thermalize quickly and
produce a thermal bath with a maximum temperature $T_{\rm max}$.
Subsequently, as inflaton decays continue, the temperature
falls with the cosmological scale factor, but not as $T \sim a^{-1}$ as is
common for an adiabatically expanding universe. Instead, the temperature
decreases more slowly, $T \sim a^{-3/8}$, as new particles are introduced
into the thermal bath from continuing decays. The reheat temperature 
is often defined when the energy density in the newly created 
radiation bath is equal to the energy density of the inflaton oscillations. 

In \cite{Bernal:2019mhf,Garcia:2020eof}, it was noted that
the evolution of the thermal bath depends on 
the form of the potential leading to inflaton oscillations.
For example, in  a class of inflation models based on attractor solutions known as `T' models \cite{Kallosh:2013hoa}, the potential in the vicinity of the minimum takes the form $V \sim \phi^k$, rather than simply $V \sim m^2 \phi^2$. In this case, it was found \cite{Garcia:2020eof} that since the effective
mass of the inflaton is now field dependent, its decay rate is as well,
thus affecting the evolution of the temperature so that $T \sim a^{-(3k-3)/(2k+4)}$. The maximum temperature as well as the reheat temperature are also affected. Here, we will show that the evolution of the temperature depends not only on $k$,
but also on the spin statistics of the final state particles produced during reheating.

The evolution of the temperature may directly affect the production of dark matter after inflation. While weakly interacting dark matter candidates will come into full equilibrium for sufficiently high 
reheat temperatures, superweakly interacting candidates
such as the gravitino \cite{gravitino,ehnos,kl,oss,nos} may be produced but never achieve thermal equilibrium before the expansion of the Universe (given by the Hubble parameter, $H$) dominates over their production rate. The same mechanism, now generally referred to as freeze-in, applies to a wider class of dark matter candidates known
as feebly interacting massive particles or FIMPs \cite{fimp,Bernal:2020gzm,Bernal,Bernal:2017kxu,Bernal:2019mhf}. Depending on the temperature 
dependence of its production rate, the relic density of a FIMP
may depend on the either the maximum temperature achieved after inflation, $T_{\rm max}$, 
the reheat temperature, $T_{\rm RH}$, or both. For example, the gravitino in weak scale
supersymmetric models is primarily dependent on the reheat temperature \cite{nos,ehnos,kl,oss,ekn,Juszkiewicz:gg,mmy,Kawasaki:1994af,Moroi:1995fs,enor,Giudice:1999am,bbb,cefo,kmy,stef,Pradler:2006qh,ps2,rs,kkmy,egnop,Garcia:2017tuj,Eberl:2020fml},
whereas the gravitino in high-scale supersymmetric models \cite{Benakli:2017whb,grav2,grav3}, depends on both $T_{\rm max}$ and $T_{\rm RH}$. 
Other examples include dark matter particles produced by 
the exchange of a massive $Z'$ \cite{Bhattacharyya:2018evo} (that can be present in SO(10) constructions \cite{SO10}) or models with a moduli portal \cite{Chowdhury:2018tzw} in emergent/modified gravity \cite{Anastasopoulos:2020gbu}. 
Even massive spin-2 \cite{Bernal:2018qlk} or
Kaluza Klein \cite{Bernal:2020fvw} fields can play the role of
an effective portal to avoid overabundance.

In all of these constructions, which can be called UV freeze-in \cite{Chen:2017kvz,Bernal:2019mhf}, 
the importance of the evolution of the temperature during reheating is extremely important.  It was shown in \cite{Garcia:2017tuj} that a large enhancement in the relic density is expected
for models whose dark matter production cross-sections are of the form
$\langle \sigma v \rangle \propto \frac{T^n}{\Lambda^{n+2}}$, with 
$n\geq 6$. While $n=0$ for the production cross section for weak scale gravitinos, $n=6$ in high-scale supersymmetric models, and $n=4$ in other spin-$\frac32$ 
dark matter models \cite{gmov}. The evolution of the temperature 
during reheating also plays a role for dark matter produced directly from 
inflaton decays, either at the tree level \cite{egnop,Garcia:2017tuj,grav2,Garcia:2020eof}, or at the loop level \cite{Kaneta:2019zgw}. 

Noting the importance of inflaton decay on the abundance of dark matter in these models,
it should not be a surprise that the shape of potential during inflaton oscillations also plays a role \cite{Bernal:2019mhf,Garcia:2020eof}. A potential of the form $V \sim \phi^k$, affects not only the equation of state and the hence the expansion
rate of the universe, but also the decay rate of the inflaton 
which becomes field dependent for $k > 2$. In this paper, we extend the recent
work of \cite{Garcia:2020eof} and show further that in models with $k \ne 2$,
the evolution of the reheating process also depends on the statistics of the
final state particles predominantly produced during inflaton decay. Furthermore,
the masses of the final state particles may also be field dependent
leading to kinematic suppressions or enhancements. Below we derive the
temperature dependence of the scale factor, the maximum temperature, and the reheat
temperature, for generic models with $k \ge 2$, and inflaton decays
into fermion/anti-fermion pairs, and boson pairs, as well as inflaton 
annihilations into boson pairs.  We also derive the thermally produced 
dark matter abundance and provide an example based on weak scale supersymmetry.

The paper is organized as follows: In section \ref{sec:osc},
we consider the effect of a potential of the form $V\sim \phi^k$ on the equation of 
state and the equations of motion governing inflaton oscillations. In section \ref{sec:decay}, we consider the decay of the inflaton to fermion and boson pairs
as well as annihilations to boson pairs. The kinematic details of these rates
are derived in the Appendix. The coupling of the inflaton leads to field 
dependent final state masses which in turn leads to a suppression in the decays to fermions,
and an enhancement in the decays to bosons. In section \ref{sec:radreh}, we
work out the general solutions for the temperature evolution during the reheat 
process and derive $T_{\rm max}$ and $T_{\rm RH}$. These results are used to 
compute the thermal production of dark matter in section \ref{sec:dm}.
A concrete example based on weak scale supersymmetry is given in section \ref{sec:susy}. Finally in section \ref{sec:disc}, we discuss the limitations of our work and summarize our results.

%%%%%%%%%%%%%%%%%%%%%%%%%%%
\section{Post-Inflationary Inflaton Oscillations}
\label{sec:osc}
%%%%%%%%%%%%%%%%%%%%%%%%%%%

In most models of inflation, the period of exponential expansion
is followed by a period of inflaton oscillations about a minimum.
These oscillations continue until the inflaton decays, and the reheating process begins 
\cite{dl}. 
The perturbative reheating approximation, fundamental for the description of post-inflationary dynamics in the small coupling limit, more often than not relies on the assumption that the inflaton is a massive field governed by the dynamics of a  quadratic potential, that is $V(\phi)\simeq \frac{1}{2}m_{\phi}^2\phi^2$ about the minimum which we assume is situated at the origin. 
If one, for simplicity, assumes that the decay of the inflaton proceeds through fermion production, $\phi\rightarrow \bar{f}f$, then its decay rate can simply be parametrized as
\beq
\Gamma_{\phi} \;\equiv\; \frac{y^2}{8\pi}m_{\phi}\,,
\eeq
where $y$ denotes the effective Yukawa coupling that determines the strength of the decay. This decay rate is a constant number, up to the running of $y$, and leads to the exponential decay of the inflaton field. Under the assumption that the decay products of $\phi$ are relativistic at their creation, and thermalize on a time scale much shorter than $\Gamma_{\phi}^{-1}$, they form a thermal bath that eventually leads to a universe dominated by radiation following the complete depletion of the energy density of $\phi$. The maximum temperature of this plasma {\em after} the decay of the inflaton is referred to as the reheating temperature, and is generically parametrized as
\beq
T_{\rm RH} \;=\; \left(\frac{40}{g_{\rm RH}\pi^2}\right)^{1/4}\left(\frac{\Gamma_{\phi}M_P}{c}\right)^{1/2}\,.
\eeq
Here, $g_{\rm RH}$ denotes the effective number of relativistic degrees of freedom at reheating time, and $c$ is an $\mathcal{O}(1)$ constant whose value depends on the convention chosen to define the reheating time. For example, $c \simeq 1$ if one assumes $t_{\rm RH}=\frac{3}{2} H(T_{\rm RH}) =\Gamma_{\phi}^{-1}$, and $c\simeq 5/3$ if instead $\rho_{\phi}(t_{\rm RH})=\rho_{r}(t_{\rm RH})$, where $\rho_{\phi}$ and $\rho_R$ denote the energy densities of the inflaton and its decay products, respectively.

As a first approximation, a quadratic potential seems natural and
is a feature of many cosmological inflationary models, among them the Starobinsky model~\cite{Starobinsky:1980te,Mukhanov:1981xt,Starobinsky:1983zz}. However, other inflationary models do not share this feature. Most notably, some $\alpha$-attractor models have minima about which $V(\phi)\sim \phi^k$ for even $k$, e.g.~the T-models~\cite{Kallosh:2013hoa}
\beq\label{eq:attractor}
V(\phi) \;=\; \lambda M^4 \left[ \sqrt{6} \tanh \left(\frac{\phi}{\sqrt{6}M}\right) \right]^k\, ,
\eeq
which can be easily derived in no-scale models of supergravity \cite{Garcia:2020eof}. Here $M$ is a characteristic mass scale of the model in question, which without loss of generality we take to be $M=M_P$, where the reduced Planck is $M_P=1/\sqrt{8\pi\,G}\simeq 2.4\times 10^{18}\,{\rm GeV}$. A potential with $k>2$ will lead to anharmonic oscillations of the inflaton field during reheating. As we discuss in more detail below, this anharmonicity is reflected in the fact that the energy density of the inflaton no longer redshifts as matter. For example, for $k=4$, $\rho_{\phi}$ redshifts like radiation, modifying drastically the reheating process when compared to the vanilla $k=2$ scenario. 
Moreover, for $k=4$,  the tree-level vacuum fluctuation of the inflaton would be massless, and its direct decay would be impossible. Nevertheless, the decay of the oscillating inflaton {\em condensate} is possible, and in the adiabatic limit it can be described by the decay of a scalar field with a time-dependent effective mass as was recently shown in \cite{Kainulainen:2016vzv,Garcia:2020eof}. Another remarkable consequence of this fact is the different time-dependence of the effective decay rate depending on the quantum statistics of the inflaton decay products. 

Furthermore, the production of dark matter during reheating will be affected by these considerations, especially in models where the production rate is highly dependent on the energy (in other words on the temperature $T$) of the scattering particles. In this work, we analyze in detail the consequences of non-quadratic inflaton-potential on the reheating processes as well as the perturbative dark matter production at the end of inflation.

We begin by considering the inflaton potential given in Eq.~(\ref{eq:attractor}).
About the origin, the potential can be expanded to give 
\beq\label{eq:Vphi}
V(\phi) \;=\; \lambda\frac{\phi^k}{M_P^{k-4}}\,, \qquad \phi \ll M_P\,.
\eeq 
Of course other inflationary potentials can be expanded about their minimum
to give a similar form as that in Eq.~(\ref{eq:Vphi}).
For example, in Starobinsky inflation, we have $k=2$
as the inflaton has a well defined mass.  
After the
exponential expansion associated with inflation, and 
during reheating, the inflaton will undergo damped oscillations about $\phi=0$. Ignoring decay for now,
the equation of motion for $\phi$ is
\beq\label{eq:phieom}
\ddot{\phi} + 3H\dot{\phi} + V'(\phi) \;=\;0 \, ,
\eeq
which in terms of the energy density and pressure stored in the scalar field
\beq
\rho_\phi = \frac{1}{2} \dot \phi^2 + V(\phi);
~~~P_{\phi} = \frac{1}{2} \dot \phi^2 - V(\phi) \, ,
\label{Eq:rhophi0}
\eeq
can be written as 
\beq
\dot \rho_\phi + 3 H(\rho_\phi + P_\phi) =0 \, ,
\label{Eq:conservation}
\eeq
where $H=\frac{\dot a}{a}$ is the Hubble parameter, and $a$ is the cosmological scale factor. 

The time dependence of the inflaton after inflation, is given by the  solution
of (\ref{eq:phieom}) and can be approximately parametrized as 
\beq
\phi(t) = \phi_0(t)\cdot \mathcal{P}(t)\,,
\nonumber
\eeq
where the function $\mathcal{P}(t)$ is quasi-periodic and encodes the (an)harmonicity of the short time-scale oscillations in the potential.
The envelope $\phi_0(t)$ encodes the effect of redshift and decay, and varies on longer time-scales.

When we include the effects of inflaton decay, the equation of motion for $\phi$ can be written as
\beq
\ddot{\phi} + (3H+\Gamma_{\phi})\dot{\phi} + V'(\phi) \;=\;0 \, .
\label{eq:phieombis}
\eeq
Provided that we assume that the decay of the inflaton is relatively slow, i.e.~the oscillation time-scale is much shorter than the decay and redshift time-scales, multiplication of (\ref{eq:phieom}) by $\phi$ and averaging over one oscillation leads to
\beq
\langle \dot{\phi}^2\rangle \;\simeq\; \langle \phi V'(\phi)\rangle\,.
\eeq
For a potential of the form (\ref{eq:Vphi}), this implies that
\bea
&&
\rho_{\phi} \simeq \frac{1}{2}\langle \dot{\phi}^2\rangle + \langle V(\phi)\rangle \simeq 
\frac{k+2}{2} \langle  V(\phi)\rangle  = V(\phi_0) \, ,
\label{Eq:rhophi2}
\\
&&
P_{\phi} \simeq \frac{1}{2}\langle \dot{\phi}^2\rangle - \langle V(\phi)\rangle \simeq 
\frac{k-2}{2} \langle V(\phi)\rangle = \frac{k-2}{k+2} V(\phi_0) \, ,
\label{Eq:Pphi}
\eea
where we used $\langle {\cal P}^k \rangle = \frac{2}{k+2}$ so that  
$\langle V(\phi) \rangle = \frac{2}{k+2}  V(\phi_0)$.
The equation of motion (\ref{eq:phieom}) can then be recast as
\beq\label{eq:rhoeom1}
\dot{\rho}_{\phi} + 3H(1+w_{\phi})\rho_{\phi} \;\simeq\; -\Gamma_{\phi}(1+w_{\phi})\rho_{\phi}\,,
\eeq
where the equation-of-state parameter $w_{\phi} = \frac{P_\phi}{\rho_\phi}$ is given by
\beq\label{eq:wk}
w_{\phi} \;=\; \frac{k-2}{k+2}\,.
\eeq
The analogous equation for the evolution of the radiation density produced by inflaton decay or scattering
(which we assume is in thermal equilibrium)
is
\beq
\dot{\rho}_{R} + 4H \rho_{R} \;\simeq\; (1+w_{\phi})\Gamma_{\phi}(t)\rho_{\phi}\, , 
\label{eq:aeom}
\eeq
which together with the Friedmann equation
\beq
\rho_{\phi}+\rho_{R} \;=\; 3H^2 M_P^2\,,
\label{hub}
\eeq 
allows one to solve for $\rho_\phi(t), \rho_R(t)$, and $a(t)$ simultaneously and effectively for $\rho_\phi(a)$
and $\rho_R(a)$. 
Comparing \mg{(\ref{eq:phieombis})} and (\ref{eq:rhoeom1}) we note that the dissipation rate from particle production for the inflaton field and energy densities differ by the constant factor 
$1+ w_\phi = \frac{2k}{k+2}$ ~\cite{Turner:1983he,Martin:2010kz}. The rate of decay for $\phi$ (and thus the number density
$n_\phi$) is different from the rate of decay for $\rho_\phi$, which depends on the nature of the inflaton field (dust, radiation, cosmological constant, quintessence...). For a microscopic account of this difference we refer the interested reader to Appendix~\ref{sec:alt}. To solve the equation for $\rho_\phi$, we must first determine the expression of the width $\Gamma_\phi$ as a function of $\phi$.

\section{Inflaton Decay and Annihilation}
\label{sec:decay}

Once the inflaton couples to Standard Model fields or dark matter, 
its oscillations are severely damped by decays.
To stay as general as possible, we consider the following possible 
contributions to the 
Lagrangian leading to decay or annihilation:
\beq
\mathcal{L}\supset  \begin{cases} 
y \phi \bar{f}f & \phi \to \bar{f}f \\
\mu \phi bb & \phi \to b b \\
\sigma \phi^2b^2 &  \phi \phi \to b  b ,
\end{cases}
\eeq
with $f\,(b$) standing for a fermionic (bosonic) final state. 
The Yukawa-like coupling, $y$ and the four-point coupling, $\sigma$, are dimensionless, and $\mu$ is a dimensionful coupling. We note that, although our analysis will be limited to these three scenarios, our formalism can extended for more exotic inflaton-matter couplings in a relatively straightforward way.

\begin{figure}[!t]
\centering
    \includegraphics[width=0.6\textwidth]{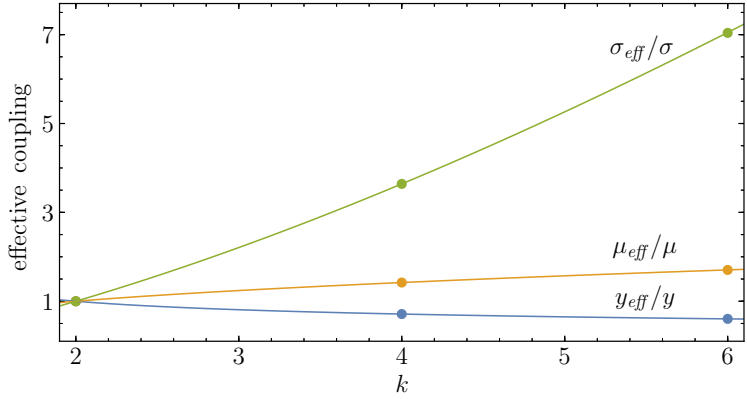}
    \caption{Numerical value of the effective inflaton matter-couplings $y_{\eff}$, $\mu_{\eff}$ and $\sigma_{\eff}$ normalized to their Lagrangian values $y$, $\mu$ and $\sigma$ respectively,
    as function of $k$. Here $m_{\eff}=0$.}
    \label{fig:couplings}
\end{figure}

Let us consider first the decay channel into two fermions. 
The rate is given by 
\beq
\Gamma_{\phi\rightarrow \bar{f}f}(t)  
\equiv\ \frac{y_{\eff}^2(k)}{8\pi}  m_{\phi}(t) ~,
\label{eq:gammaff}
\eeq
where we have introduced the effective Yukawa coupling $y_{\eff}(k) \neq y$ obtained after averaging over one oscillation, and $m_\phi$ is defined by
\beq
\label{eq:mephiff}
m_{\phi}^2(t) \;\equiv\; V''(\phi_0(t)) \;=\; k(k-1)\lambda
M_P^2 \left(\frac{\phi_0(t)}{M_P}\right)^{k-2}\,.
\eeq
The function $y_{\eff}(k)$ includes sub-leading corrections, and must be evaluated numerically~\cite{Shtanov:1994ce,Ichikawa:2008ne}. It is different from
$y$ because for $k\neq 2$, 
the inflaton mass depends on the oscillations of the field $\phi(t)$ 
and renders the lifetime computation slightly more complicated, and must include the 
mean of several oscillations. 
In (\ref{eq:gammaff}) the time dependence of
$m_\phi(t)$ is included in the envelope $\phi_0(t)$ only,
which will be our main dynamical parameter during all our analysis.
Note that this is analogous to $\rho_\phi = V(\phi_0)$ (\ref{Eq:rhophi2}) which is defined
as function of the envelope. This can be understood by noticing that at the top of an oscillation, $\dot \phi$ is zero, and the inflaton 
behaves like a massive particle at rest.  
 However, for the curious reader, we derive $y_{\eff}$ in appendix \ref{sec:alt}, 
Eq.~(\ref{eq:yeffdef}), and we show the result of our numerical calculation $\frac{y_{\eff}}{y}$ in Fig.~\ref{fig:couplings}.
For $k=2$, $y_{\eff} = y$, since  the oscillations obviously do not affect the inflaton mass. For $k = 6$, we find a reduction in the coupling by approximately 40\%. For simplicity, we will write from now on $y_{\eff}$ for $y_{\eff}(k)$.

When the inflaton decays into a pair of scalars, the decay rate takes the form
\beq
\Gamma_{\phi\rightarrow bb}(t) \equiv 
\frac{\mu_{\eff}^2(k)}{8\pi  m_{\phi}(t) }~,
\label{eq:gammapbb}
\eeq
where $\mu_{\eff}(k)$ is a weakly-dependent function of $k$, shown in Fig.~\ref{fig:couplings}, where, as discussed for $y_{\eff}$, for $k=2$, $\mu_{\eff} = \mu$ and the largest variation is also no larger than a factor of 1.7. 
The exact expression of $\mu_{\eff}(k)$ as function of the Lagrangian parameter $\mu$ can also be found in appendix \ref{sec:alt}, Eq.~(\ref{eq:zeffdef}).
Finally, if we consider the four-point process, the time-dependent dissipation rate will be given by\footnote{Note that for $k=2$, the rate can be written in the familiar form: $\Gamma_{\phi \phi \rightarrow b b} = n_\phi(t) \langle\sigma v\rangle_{\phi \phi \rightarrow bb} = \frac{\rho_\phi(t)}{m_\phi(t)} \frac{|{\cal M }|^2}{16 \pi m_\phi^2}$, where ${\cal M}$ is the scattering amplitude of the process $\phi \phi \rightarrow b b$.}
\beq
\Gamma_{\phi \phi \rightarrow b b}  = \frac{\sigma_{\eff}^2}{8 \pi} 
\frac{\rho_\phi(t)}{ m^3_\phi(t)}. 
\label{eq:gammappbb}
\eeq
The sub-leading correction is shown in Fig.~\ref{fig:couplings}, and the
analytical expression for $\sigma_{\eff}(k)$ as a function of $\sigma$
is given by Eq.~(\ref{eq:seffdef}).
The normalization for the decay rate is chosen so that $\sigma_{\eff} = \sigma$ for $k=2$. 
From the expressions above, we understand clearly how the shape 
of the inflaton potential will influence its decay rate through its mass
$m_\phi(t)$ (\ref{eq:mephiff}) and its density $\rho_\phi(t)$ (\ref{Eq:rhophi0}) which becomes $k$-dependent.

Before going into the details of the analysis, we can attempt to understand
the behavior of inflaton decay by looking at its width.
The decay into fermions is proportional to $m_\phi(t)$ and thus to
$\phi_0(t)^{\frac{k-2}{2}}$, whereas the decay into bosons is proportional to 
$\frac{1}{m_\phi(t)}$ {\it i.e.}~$\phi_0(t)^{\frac{2-k}{2}}$. We see
then that the reheating process will be more efficient over time for bosonic
final states than fermionic final states, because $\phi_0(t)$ is a decreasing function of time (for $k$ larger than 2). We then expect a steeper slope for the temperature $T$ as a function of the scale factor for the fermions than for bosons in the final state (roughly speaking, larger decay rates means larger temperature).
In further contrast, for the $\phi \phi \rightarrow bb$ process, the width will be proportional to $\Gamma_{\phi \phi \rightarrow bb} \propto \phi_0^{3-\frac{k}{2}}(t)$, which means that it is always more efficient than 
$\phi \rightarrow bb$ process over time and less efficient than $\phi \rightarrow \bar{f}f$ process for $k \leq 4$, modulo the relative value of the couplings of course. These features are summarized in Table \ref{Tab:table}
that will be explained in due course. 
The value of the field $\phi(t)$, acting as a background field,
also generates dynamical masses to 
the final products $f$ and $b$, which 
therefore depends on shape of the inflaton potential, opening the possibility
of dynamic kinematic blocking during the reheating phase.

The rates for the inflaton decay processes that we have introduced above, namely (\ref{eq:gammaff}), (\ref{eq:gammapbb}) and (\ref{eq:gammappbb}), implicitly assume that the decay products of the inflaton are massless. However, as the oscillations of the inflaton provide a background in which $\phi$ acquires an effective mass, the same will occur for the decay products $f$ and $b$. The tree-level couplings of these fields to the inflaton lead to the following form for their time-dependent effective masses,
\beq\label{eq:mprod}
m_{\eff}^2(t) \;\equiv\; 
\begin{cases}
y^2 \phi^2
\,, & \phi\rightarrow \bar{f}f\,,\\
2\mu \phi
\,, & \phi\rightarrow bb\,,\\
2\sigma \phi^2
\,, & \phi\phi\rightarrow bb\,.
\end{cases}
\eeq
Hence, the condition $m_{\eff}^2(t) \ll m_{\phi}^2$ for the efficient population of the {\em relativistic} plasma from inflaton decay is in general a time-dependent statement. At the perturbative level, disregarding the short time-scale of oscillations of $\phi$, the effect of this time-dependent effective mass can be determined upon averaging over the oscillations the effective decay rate. This procedure is discussed in detail in Appendix~\ref{sec:alt}. The parameter which determines the relevance of the induced mass is given by
\beq\label{eq:kincond}
\mathcal{R}(t) \;\equiv\; \frac{8}{\pi k^2 \lambda}\left(\frac{\Gamma(\frac{1}{k})}{\Gamma(\frac{1}{2}+\frac{1}{k})}\right)^2 \times \begin{cases}
y^2 \left(\dfrac{\phi_0(t)}{M_P}\right)^{4-k}\,, & \phi\rightarrow \bar{f}f\,,\\[9pt]
2\dfrac{\mu}{M_P} \left(\dfrac{\phi_0(t)}{M_P}\right)^{3-k}\,, & \phi\rightarrow bb\,,\\[9pt]
2\sigma \left(\dfrac{\phi_0(t)}{M_P}\right)^{4-k}\,, & \phi\phi\rightarrow bb\,.
\end{cases}
\eeq
Note that $\mathcal{R}\propto (m_{\eff}/m_{\phi})^2|_{\phi\rightarrow \phi_0}$. For $\mathcal{R}\ll 1$, the effective mass of the decay products is much smaller than the inflaton mass, and any kinematic effects can be safely disregarded. However, for $\mathcal{R}\gtrsim 1$, the phase-space dependence on $m_{\eff}$ must be taken into account. For the $\phi\rightarrow \bar{f}f$ and $\phi\phi\rightarrow bb$ cases, for which $m_{\eff}\propto \phi^2$, the result is a suppression of the mean decay rate of $\phi$. This kinematic blocking is, however, not total, as for any value of the coupling there will exist a time interval around the moment when $\phi=0$ during which the decay is allowed. A numerical evaluation of the corresponding phase-space factors reveals that $\Gamma_{\phi} \propto \mathcal{R}^{-1/2}$ when $\mathcal{R}\gg 1$ (see Appendix~\ref{sec:alt}). On the other hand, for $\phi\rightarrow bb$, $m_{\eff}\propto \phi$, and hence for half of the inflaton oscillation this effective mass becomes negative. Therefore on average not only there is no kinematic suppression in this scenario for $\mathcal{R}\gg 1$, but in fact there is an enhancement of the decay rate, $\Gamma_{\phi} \propto \mathcal{R}^{1/2}$ (see Fig.~\ref{fig:KinB1}). This steep enhancement of the inflaton width is related to the tachyonic excitation of $b$, which signals the breakdown of the perturbative approximation and the need to consider the short-time preheating effects. In fact, the condition $\mathcal{R}\gg 1$ coincides, in all three cases, with the broad resonance regime, in which the non-perturbative production of {\em non-relativistic} decay products can be efficient~\cite{Kofman:1997yn,Greene:1997fu}. We will not consider this case in our work, and will be the subject of an independent analysis.

As one can see from (\ref{eq:kincond}), depending on the value of $k$ and the primary mode for decay (or scattering),
the ratios in (\ref{eq:kincond}) will scale as $\phi_0^p$
where $p$ may be positive or negative.  As we are
implicitly assuming that the inflaton is evolving from
an initially large value to the origin, the ratios in (\ref{eq:kincond}) may either increase or decrease. 
Consider for example that inflaton decay has a dominant fermionic decay channel (or if the depletion of the inflaton is dominated by $\phi\phi\rightarrow bb$). In this case $\mathcal{R}$ {\it decreases} for $k<4$. Therefore, if inflaton decay is not kinematically suppressed when inflation ends at $t = t_{\rm end}$, it will also not be suppressed at any subsequent time. If at $t = t_{\rm end}$ the decay is suppressed, the efficient decay of $\phi$ is delayed until $\phi_0(t)$ decreases sufficiently to allow the decay. For $k>4$, the mass ratio (\ref{eq:kincond}) increases with decreasing $\phi_0$ in these channels and even if decay is possible at $t = t_{\rm end}$, it becomes blocked at later times. For boson dominated decays, these qualitative effects
depend on $p > {\rm or} < 3$. However, in this case instead of a decrease in the efficiency of the decay, we observe the breakdown of the perturbative approximation due to an increase in the rate. We comment on this further in the discussion section. 

\begin{figure}[!ht]
\centering
    \includegraphics[width=\textwidth]{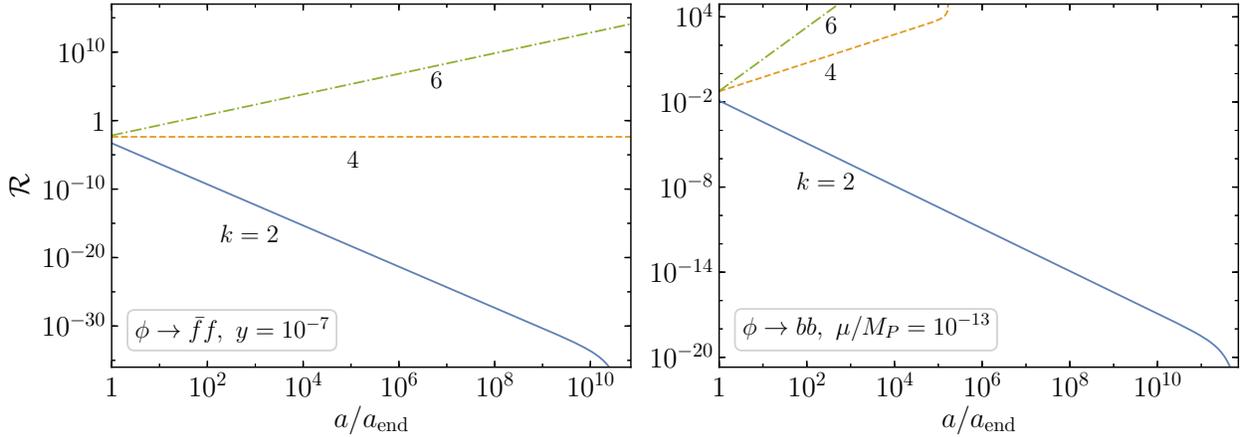}
    \caption{The kinematic parameter $\mathcal{R}$ defined in (\ref{eq:kincond}) for fermionic (left) and bosonic (right) decays of the inflaton, as a function of the scale factor for $k=2,4,6$. T-attractor values are chosen for the potential parameter $\lambda$ and the inflaton $\phi_{\rm end}$ (see Appendix~\ref{sec:efold}). Inflaton-radiation equality occurs at $a_{\rm RH}/a_{\rm end}\simeq 10^{10}\,(2\times 10^{11}),\,7\times 10^{14}\,(10^5)$ and $2\times 10^{28}\,(4\times 10^3)$ for fermions (bosons) with $k=2,4,6$, respectively.}
    \label{fig:kinsup}
\end{figure}

Fig.~\ref{fig:kinsup} shows the evolution of the mass ratio $\mathcal{R}$ for the case of inflaton decays into fermions and bosons, for $k=2,4,6$ as function of $\frac{a}{a_{\mathrm{end}}}$,
where $a_{\rm end} = a(t_{\rm end})$ denotes the scale factor at the end of inflation. For fermions, shown in the left panel, we find, as expected, that $\mathcal{R}$ decreases for $k<4$, is constant for $k=4$ and increases for $k>4$. For the chosen value of $y$, the effect of the kinematic suppression for $k\leq 4$ can be neglected. However, for $k=6$, the decay quickly becomes kinematically blocked, resulting in a reduced decay rate. This reflects the fact that for large values of $k$, the inflaton mass (\ref{eq:mephiff}) redshifts faster than that of the decay product masses (\ref{eq:mprod}) which are independent of $k$. 
For reference, the delay of inflaton-radiation equality would lead to a reheating temperature $T_{\rm RH}\simeq 4\times 10^{-17}\,{\rm GeV}$, lower than the present photon temperature and in clear conflict with cosmological constraints including big bang nucleosynthesis~\cite{Hasegawa:2019jsa,Fields:2019pfx}. On the other hand, for the bosonic decay channel, we observe that $\mathcal{R}$ decreases for $k=2$, but it rapidly increases for $k=4,6$. This results in a shortened reheating epoch. We can then conclude that the shape of the inflationary potential about the origin as determined by the value of $k$ has a strong effect on the kinematics of the final state, in addition to its effect on the inflaton width. We are now in a position to analyze the evolution of the temperature of the thermal bath produced by inflaton decay and scattering.%

\section{The Reheating Process}\label{sec:radreh}

We use Eqs.~(\ref{eq:rhoeom1}), (\ref{eq:aeom}), and (\ref{hub}) to determine the time evolution of the energy density of the decay products of the inflaton during reheating.
We can write the dissipation rate in terms of $\rho_{\phi}$ to obtain a closed set of evolution equations. If we 
average over several oscillations and combine Eqs.~(\ref{eq:gammaff}),  (\ref{eq:mephiff}), (\ref{eq:gammapbb})
and (\ref{eq:gammappbb}), we have
\beq
\Gamma_{\phi}(t) \;=\;
\gamma_{\phi}\left(\frac{\rho_{\phi}}{M_P^4}\right)^{l}\,,
\label{Eq:gammaphi}
\eeq
where
\beq
\label{Eq:gammaphibis}
\gamma_{\phi} \;=\; 
\begin{cases}
\sqrt{k(k-1)}\lambda^{1/k}M_P\dfrac{y_{\eff }^2}{8\pi}\,,\quad & \phi\rightarrow \bar{f}f\,,\\[10pt]
\dfrac{\mu_{\eff }^2}{8\pi\sqrt{k(k-1)}\lambda^{1/k}M_P}\,,\quad& \phi\rightarrow bb\,,\\[10pt]
\dfrac{\sigma_{\eff}^2 M_P}{8\pi [k(k-1)]^{3/2}\lambda^{3/k}}\,,\quad & \phi\phi\rightarrow bb\,,
\end{cases}
\eeq
and
\beq
l \;=\; 
\begin{cases}
\frac{1}{2}-\frac{1}{k}\,,\quad & \phi\rightarrow \bar{f}f\,,\\
\frac{1}{k}-\frac{1}{2}\,,\quad & \phi\rightarrow bb\,,\\
\frac{3}{k}-\frac{1}{2}\,,\quad & \phi\phi\rightarrow bb\,.
\end{cases}
\label{ls}
\eeq
Multiplying both sides of Eq.~(\ref{eq:rhoeom1}) by 
$a^{\frac{6k}{k+2}}$, replacing $\Gamma_\phi$ by (\ref{Eq:gammaphi}),
replacing $w_\phi$ using Eq.~(\ref{eq:wk}), 
and replacing the dynamical parameter $t$ by $a$ with 
$\frac{d}{dt} = a H \frac{d}{da}$ we obtain
\beq
\frac{d}{da}\left( \rho_\phi a^{\frac{6k}{k+2}} \right) =
- \frac{\gamma_\phi}{aH} \frac{2k}{k+2} \frac{\rho_\phi^{l+1}}{M_P^{4l}}
a^{\frac{6k}{k+2}} \, .
\eeq
If we now suppose $\gamma_\phi \ll H$, valid at early times,
\beq
\rho_\phi(a) = \rho_{\rm end} \left(\frac{a}{a_{\rm end}} \right)^{-\frac{6k}{k+2}} \, ,
\label{Eq:rhophi}
\eeq
where $\rho_{\rm end} = \rho_\phi(a_{\rm end})$.
At later times, for $k=2$, $\rho_{\phi}\propto e^{-\Gamma_{\phi}t}$, however, the decay of the inflaton is {\em not} exponential for $k>2$.

Inserting Eq.~(\ref{Eq:rhophi}) into (\ref{eq:aeom}) and performing a similar manipulation,
we have
\beq
\frac{1}{a^4} \frac{d }{da}\left( \rho_R a^4\right) \;=\; \frac{2k}{k+2}
\frac{\gamma_\phi}{aH}
\frac{\rho_\phi^{l+1}}{M_P^{4l}} \, .
\label{diffa}
\eeq
This expression is easily integrated to give
\beq
\rho_R \; = \; \frac{2 k}{k + 8 - 6kl} \frac{\gamma_\phi}{H_{\rm end}}
 \frac{\rho_{\rm end}^{l+1}}{M_P^{4l}} \left( \frac{a_{\rm end}}{a} \right)^4
\left[ \left( \frac{a}{a_{\rm end}}\right)^{\frac{k+8-6kl}{k+2}} -1\right] \, ,
\label{Eq:rhoR}
\eeq
where $H^2_{\rm end} = \rho_{\rm end}/3 M_P^2$. 
At later times when $a \gg a_{\rm end}$ and $8+k-6kl>0$,
we can approximate $\rho_R$ as
\beq
\rho_R^{a\gg a_{\rm end}} \; = \; \frac{2 k}{k + 8 - 6kl} \frac{\gamma_\phi}{H_{\rm end}}
 \frac{\rho_{\rm end}^{l+1}}{M_P^{4l}} \left( \frac{a_{\rm end}}{a} \right)^{\frac{3k + 6kl}{k+2}} \, .
\label{Eq:rhoRapp}
\eeq
For the case with dominant inflaton decays to fermions,  $\phi\to f\bar f$, when $l=(k-2)/2k$, we recover the result in Ref.~\cite{Garcia:2020eof} .

Given the expression for $\rho_R$ in Eq.~(\ref{Eq:rhoR}), the temperature of the radiation bath
is simply
\beq
\rho_R = \frac{g_\rho \pi^2}{30} T^4 ~~~~\Rightarrow~~~~
T = \left( \frac{30 \rho_R}{g_\rho \pi^2} \right)^{\frac{1}{4}} 
\propto a^{-\frac{3k+6kl}{4k+8}} \, ,
\label{Eq:tfa}
\eeq
where
$g_{\rho}$ is the number of relativistic degrees of freedom at temperature $T$. Note that if $8+k-6kl<0$,
\beq
\rho_R^{a\gg a_{\rm end}} \;=\; \frac{2k}{6kl-k-8} \frac{\gamma_\phi}{H_{\rm end}} \frac{\rho_{\rm end}^{l+1}}{M_P^{4l}}  \left(\frac{a_{\rm end}}{a}\right)^{4}\,,
\label{Eq:rhoRappter}
\eeq
which implies that the temperature would simply redshift as $T\propto a^{-1}$.
As a summary, we provide in Table~\ref{Tab:table} the dependence of
$T$ as function of $a$ for the different cases we analyze
in our work. In the last column of the table, we show the 
form of the temperature evolution when ${\cal R} \gg 1$.

\begin{table*}[!ht]
\centering
%\begin{tabular*}{\columnwidth}{@{\extracolsep{\fill}}lllll@{}}
\bgroup
\def\arraystretch{1.5}
\begin{tabular}{|c||c|c|c|c|c|}
\hline 
channel & generic & $k=2$ & $k=4$ & $k=6$ & $m^2_{\eff}\gg m^2_\phi$  
 \\ 
 \hline \hline
$\phi\rightarrow \bar f f$ & $T\propto a^{-\frac{3k-3}{2 k + 4}}$ & $T \propto a^{-3/8}$ & $T \propto a^{-3/4}$ & $T \propto a^{-15/16}$ & $T \propto a^{-\frac{9(k-2)}{4(k+2)}}$ 
\\
 \hline 
$\phi\rightarrow bb$ & $T\propto a^{-\frac{3}{2k+4}}$ & $T \propto a^{-3/8}$ & $T \propto a^{-1/4}$ & $T \propto a^{-3/16}$ & $T \propto a^{-\frac{3(5-k)}{4(k+2)}}$ 
\\
 \hline 
$\phi \phi \rightarrow b b$ & $T\propto a^{- \frac{9}{2 k + 4}}$ & $T \propto a^{-1} $ & $T \propto a^{-3/4}$ & $T \propto a^{-9/16}$ & $T \propto a^{-3/4}$ 
\\
\hline
\end{tabular}
\egroup
\caption{Dependence of the temperature $T$ as function of the scale factor $a$ for the different cases we analyze in this work. The `generic' result assumes the validity of Eq.~(\ref{Eq:rhoRapp}). In the last column non-perturbative particle production has not been taken into account, even if $\mathcal{R}\gg 1$.}
\label{Tab:table}
\end{table*}

At the end of inflation, before inflatons decay, $\rho_R = 0$ and hence $T=0$.
The Universe begins to reheat and a maximum temperature is attained before the
temperature begins to fall off as given in Table~\ref{Tab:table}.
The maximum temperature can be computed from Eq.(\ref{Eq:rhoR}). From $\frac{d\rho_R}{da}=0$, we obtain
\beq
a_{\rm max}= a_{\rm end}\left(\frac{4k+8}{3k + 6kl} \right)^{\frac{k+2}{k+8-6kl}} \, ,
\label{maxend}
\eeq
which gives 
\beq
\rho_R^{\rm max} \;=\;  \frac{2}{3+6l} \frac{\gamma_\phi}{H_{\rm end}} \frac{\rho_{\rm end}^{l+1}}{M_P^{4l}} \left(\frac{4k + 8}{3k+6kl}\right)^{-\frac{4k+8}{k+8-6kl}} \, ,
\label{rhomax}
\eeq
and
\beq
T_{\rm max} = \left(
\frac{30}{g_\rho \pi^2} \rho_R^{\rm max} \right)^{\frac{1}{4}}.
\label{Tmax}
\eeq

\begin{figure}[!ht]
\centering
    \includegraphics[width=\textwidth]{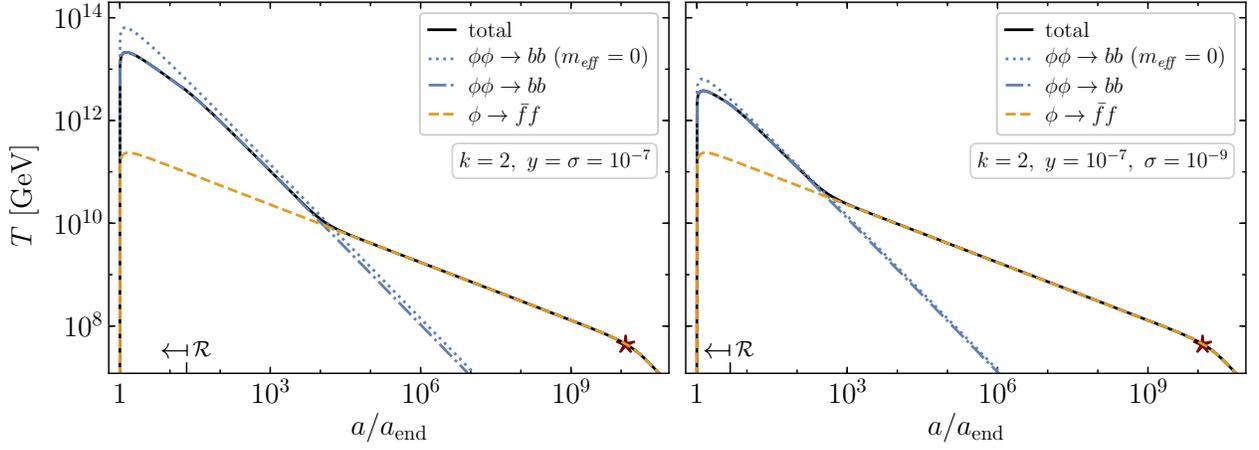}
    \caption{ Evolution of the instantaneous temperature during reheating for $k=2$ in the case of inflaton decays to fermions (dashed, orange) and annihilations to bosons (blue). In the latter, we show separately the case when the effective masses of the decay products are ignored (dotted) and included (dot dashed). The case of both decays and annihilations (with effective masses included) is also shown (solid, black). In the left panel we take $y = \sigma = 10^{-7}$ and in the right panel $y = 10^{-7}$ and $\sigma  = 10^{-9}$. Here $\rho_{\rm end}=(5.2\times 10^{15}\,{\rm GeV})^4$ and $\lambda=2.5\times 10^{-11}$, assuming T-attractor inflation boundary conditions.
    The star signals inflation-radiation equality. The arrow points toward the region where $\mathcal{R}>1$ for one or more of the decay channels.}
    \label{fig:mixedtempsk2}
\end{figure}

\begin{figure}[!ht]
\centering
  \includegraphics[width=\textwidth]{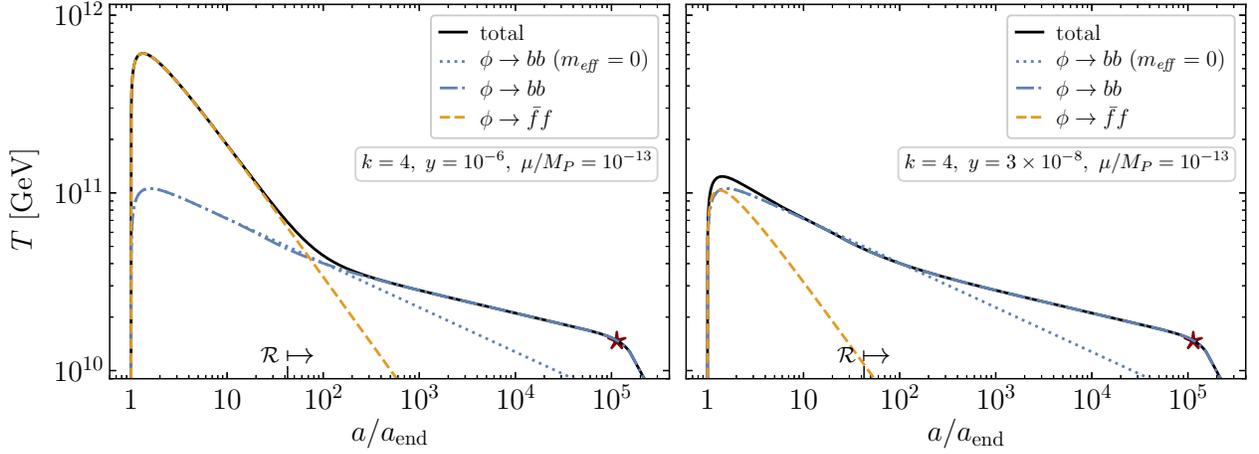}
    \caption{ Evolution of the instantaneous temperature during reheating for $k=4$ in the case of inflaton decays to fermions (dashed, orange), bosons (blue). In the latter, we show separately the case when the effective masses of the decay products are ignored (dotted) and included (dot dashed). The case of both decays and annihilations (with effective masses included) is also shown (solid, black).  In both cases the bosonic coupling is fixed to $\mu=10^{-13}\,M_P$. The left panel depicts $T$ vs.~$a$ for $y=10^{-6}$. The right panel corresponds to $y=3\times 10^{-8}$. Here $\rho_{\rm end}=(4.8\times 10^{15}\,{\rm GeV})^4$ and $\lambda=3.3\times 10^{-12}$, assuming T-attractor inflation boundary conditions. The star signals inflation-radiation equality. The arrow points toward the region where $\mathcal{R}>1$ for one or more of the decay channels.}
    \label{fig:mixedtemps1}
\end{figure}

We show in Figs.~\ref{fig:mixedtempsk2} and \ref{fig:mixedtemps1} 
the evolution of the temperature obtained by numerically solving Eqs.~(\ref{eq:rhoeom1})-(\ref{hub}), as function of the scale factor $a/a_{\rm end}$
for two choices of $k = 2$ and 4. To see the effect of the kinematic suppression, we compare the results where $m_{\eff}$ is given by Eq.~(\ref{eq:mprod}) to one where we set $m_{\eff} = 0$. 
We begin by considering the case with $k=2$.
The value $\phi_{\rm end}$ is determined by the condition that exponential 
expansion ceases, or $\ddot{a} = 0$. The scale of the potential, $\lambda$
can be obtained by the normalization of the CMB and the number of $e$-folds since 
horizon crossing.  This procedure is worked out for the T-attractor models in 
Appendix \ref{sec:efold}.  For $k=2$ we find
$\lambda=2.5 \times 10^{-11}$  and 
$\rho_{\rm end}^{1/4} = 5.2 \times 10^{15}$ GeV.
Since we expect the evolution of the temperature to be similar for the cases of decays to bosons and fermions (see Table \ref{Tab:table}), we include only decays to fermions and annihilations to boson pairs. In Fig.~\ref{fig:mixedtempsk2}, we take
$y = \sigma = 10^{-7}$ (left) and  $y = 10^{-7}$ and $\sigma  = 10^{-9}$ (right). For inflaton decays to fermions, we can estimate the maximum temperature attained from Eqs.~(\ref{rhomax}) and (\ref{Tmax}), 
\beq
\rho_R^{\rm max} = \frac{\sqrt{6}}{32 \pi} \left(\frac{3}{8}\right)^{3/5} ~y^2 M_P^2 
\left( \lambda
\rho_{\rm end}
\right)^{\frac{1}{2}} 
~~~ \Rightarrow ~~~  T_{\rm max} \sim 2 \times  10^{11} \left(\frac{y}{10^{-7}} \right)^{1/2} ~ \rm{GeV},
\eeq
in good agreement with the numerical result shown in the figure.
Similarly, for annihilations to boson pairs, 
we expect
\beq
\rho_R^{\rm max} = \frac{\sqrt{3/2}}{72 \pi M_P^2} \left(\frac{8}{9}\right)^{8} ~\sigma^2 \lambda^{-\frac{3}{2}}
\left(
\rho_{\rm end}
\right)^{\frac{3}{2}} 
~~~ \Rightarrow ~~~  T_{\rm max} \sim 6 \times  10^{12} \left(\frac{\sigma}{10^{-9}} \right)^{1/2} ~ \rm{GeV},
\eeq
 which is close to the result shown in the figure for the case where
the kinematic suppression in the final state is ignored (the dotted curves with $m_{\eff} = 0$). 

Also apparent in Fig.~\ref{fig:mixedtempsk2} is the difference in the slopes of the evolution, $T \propto a^{-3/8}$ for decays to fermions, and $T \propto a^{-1}$, for annihilations to bosons
(see again Table~\ref{Tab:table}). 
We can estimate the value of $a/a_{\rm end}$ for which the two contributions are equal by
using Eqs.~(\ref{Eq:rhoRapp}) and
(\ref{Eq:rhoRappter}), and we obtain
\beq
\rho^f_R=\rho^b_R ~\Rightarrow~~
\left(\frac{a}{a_{\rm end}} \right)
=\left(\frac{5}{4}\frac{\rho_{\rm end}}{M_P^4}\frac{\sigma^2}{y^2 \lambda^2}  \right)^{2/5}
\simeq 18000 ~~\rm{for}~ \sigma={10^{-7}}~\rm{and}~ \simeq 450 ~\rm{for} 
~\sigma=10^{-9}.
\eeq
 This is in reasonable agreement with the numerical result in Fig.~\ref{fig:mixedtempsk2}. 
For the value of $y$ adopted in Fig.~\ref{fig:mixedtempsk2}, the value of $\mathcal{R}$ in Eq.~(\ref{eq:kincond}) is much smaller than one, and we do not expect (and do not find) any kinematic suppression for the evolution of $T$ produced by decays to fermions. 
In contrast, we do find some suppression for the case of inflaton
annihilations to bosons. This is evidenced by the suppression in
$T_{\rm max}$ and the change in slope in the blue dot-dashed curve
when compared with the dotted curve for which the effect is neglected. We can estimate the value of $a$ for which the change in slope occurs from the condition $\mathcal{R} \simeq 1$. For $k=2$, we find
\beq
\mathcal{R} = \frac{4 \sigma \rho_{\rm end}}{\lambda^2 M_P^4} \left(\frac{a_{\rm end}}{a} \right)^3 \simeq 1 \ \ \Rightarrow \ \  a \simeq 20 ~\left( \sigma/10^{-7} \right)^{1/3}~a_{\rm end} \, ,
\eeq
where we have used Eq.~(\ref{Eq:rhophi}) and $\rho_\phi = V(\phi_0)$, 
Eq.~(\ref{Eq:rhophi2}). Once again, our analytic approximation is in good agreement with the position of the change in slope seen in Fig.~\ref{fig:mixedtempsk2}.  Finally, as noted earlier, the effect of the kinematic suppression 
causes an effective reduction of the decay rate by $\mathcal{R}^{-1/2}$ when $\mathcal{R}>1$. For $l = \frac{3}{k} - \frac{1}{2}$ (the value corresponding to the
process $\phi \phi \rightarrow bb$), 
$\mathcal{R}^{-1/2} \sim a^{(12-3k)/(k+2)}$ and integrating Eq.~(\ref{diffa})
with $\gamma_\phi \to \gamma_\phi \mathcal{R}^{-1/2}$, we find 
$T\sim a^{-3/4}$ for all values of $k$  when kinematic suppression is important, 
and $T\sim a^{-1}$ at later times, 
when the suppression is no longer important as seen in Fig.~\ref{fig:mixedtempsk2}.  Note also the change in slope at $\left (\frac{a}{a_{\rm end}} \right)\simeq 10^{10}$, where $T$ becomes proportional to $a^{-1}$, corresponding to the reheat temperature
$T=T_{\rm RH}$ defined by $\rho_\phi = \rho_R$
and discussed in more detail below.

We next consider the evolution of the temperature for the case with $k=4$. In this case, the evolution of the temperature due to annihilations to bosons is similar to that from decays to fermions, and we ignore inflaton annihilations by setting $\sigma = 0$. 
In Fig.~\ref{fig:mixedtemps1}, we compare the evolution of the temperature for two choices of the fermionic coupling, $y = 10^{-6}$ (left) and $y = 3\times 10^{-8}$ (right) for a common coupling to bosons, $\mu = 10^{-13} M_P$. From the normalizations derived in Appendix \ref{sec:efold}, we now find $\lambda=3.3\times 10^{-12}$ and $\rho_{\rm end}^{1/4}=4.8\times 10^{15}\,{\rm GeV}$. 
From 
Eqs.~(\ref{rhomax}) and (\ref{Tmax}), we find
\beq
\rho_R^{\rm max} = \frac{27}{256 \pi} ~y^2_{\eff}M_P \lambda^{\frac{1}{4}}
\rho_{\rm end}^{\frac{3}{4}} 
~~~ \Rightarrow ~~~  T_{\rm max} \sim 6 \times  10^{11} \left(\frac{y}{10^{-6}} \right)^{1/2} ~ \rm{GeV},
\eeq
for our assumed values of $\lambda$, and $\rho_{\rm end}$ with $g_\rho \sim 100$. This is very close to the maximum temperature attained in the numerical result shown in Fig.~\ref{fig:mixedtemps1}.  
For $y = 10^{-6}$, we see that the initial
stages of reheating are dominated by
fermionic final states, and the temperature evolution is governed by
$T \propto a^{-3/4}$ as expected from Eq.~(\ref{Eq:tfa}), 
until 
decays to bosons become important.
Decays to bosons lead to a maximum temperature 
given by
\beq
\rho_R^{\rm max} =\frac{\mu^2_{\eff}}{12 \pi}4^{-\frac{4}{3}}
\lambda^{-\frac{1}{4}} M_P\rho_{\rm end}^{\frac{1}{4}}
~~\Rightarrow ~~~T_{\rm max} \simeq 10^{11} \left(\frac{\mu}{10^{-13}M_P} \right)^{1/2} ~\rm{GeV}. 
\eeq
However the temperature produced from decays to bosons falls off slower, as $T \propto a^{-1/4}$ 
and bosonic reheating dominates when 
\beq
\frac{a}{a_{\rm end}} \;=\; 6 \frac{y_{\eff}}{\mu_{\eff}}\left( \lambda \rho_{\rm end}\right)^{1/4} \;\simeq\; 80 \, ,
\eeq
from Eq.(\ref{Eq:rhoRapp}) for $y = 10^{-6}$ and $\mu = 10^{-13} M_P$.
This corresponds to what we obtained numerically in Fig.~\ref{fig:mixedtemps1}.

For larger values of $a$, the bosonic
gas, even if less populated at the beginning of reheating, because of our
choices of  $y_{\eff}$ and $\mu_{\eff}$, begins
to dominate the energy budget of the thermal bath. This comes from the fact the whereas the production rate of fermions decreases with $\phi_0$ (Eq.~\ref{eq:gammaff}), the opposite is true for the process $\phi \rightarrow b b$ which becomes more efficient 
with time (Eq.~\ref{eq:gammapbb}). This is reflected in the temperature evolution, $T \propto a^{-3/4}$ for the fermionic plasma and $T \propto a^{-1/4}$ for a bosonic plasma (see Table~\ref{Tab:table}). On the other hand,
if we set $y_{\eff} = \frac{\mu_{\eff}}{m_\phi}=3\times 10^{-8}$
for $\mu = 10^{-13} M_P$  as illustrated in Fig.~\ref{fig:mixedtemps1} (right)\footnote{In other words same effective coupling to inflaton (compare Eqs.~\ref{eq:gammaff} and \ref{eq:gammapbb}).}, 
we will obtain roughly the same amount of fermionic and bosonic components
at $T_{\rm max}$, but because the temperature evolves differently for the two species, only the bosonic final states reheat the Universe.

The value of $a$ for which the bosonic enhancement factor
${\cal R}$ plays a significant role is given by ${\cal R} \gtrsim 1$,
or
\beq
{\cal R} = \frac{\mu}{\lambda^{3/4} \pi \rho_{\rm end}^{1/4}} 
\left[ \frac{\Gamma(\frac{1}{4})}{\Gamma(\frac{3}{4})}\right]^2 
\left( \frac{a}{a_{\rm end}}\right) \gtrsim 1 
\quad\ \Rightarrow \quad\  a \gtrsim 20\,a_{\rm end} \, .
\label{Eq:k4r1}
\eeq
The numerical result shows that the slope change occurs around $a/a_{\rm end} \sim 100$, indicating the effect of the enhancement requires ${\cal R} \sim 5$ (see Appendix~\ref{app:phitobb}). 
    At large $a$, when the enhancement is effective, the slope changes from $T \propto a^{-1/4}$ to   $T \propto a^{-1/8}$,
corresponding to the shallow slope seen in Fig.~\ref{fig:mixedtemps1}. It must be emphasized that a significant amount of uncertainty is present, since we have neglected non-perturbative particle production.

When we decrease $y$ so that the value of $T_{\rm max}$ produced 
by decays to fermions is approximately equal to that as decays to bosons as in 
Fig.~\ref{fig:mixedtemps1} (right), we observe that, as expected, 
the reheating
is first dominated by the process $\phi \rightarrow bb$.  In the absence of kinematic blocking, the temperature of the plasma due to final state bosons falls off as $T\sim a^{-1/4}$ until the end of reheating (when $t \simeq \Gamma_\phi^{-1}$). However, 
kinematic
enhancement turns on at $a \simeq 100~a_{\rm end}$ and 
the temperature falls off more gradually as $T\sim a^{-1/8}$
until the radiation bath dominates the energy density
at $a_{\rm RH} \simeq 10^5~a_{\rm end}$ which we define as the moment of reheating and subsequently $T \sim a^{-1}$
as discussed further in the next subsection.

As we have seen in the previous subsection,
reheating is a continuous process as inflaton decays products appear and thermalize. 
We define the reheat temperature when 
\beq
\rho_R(T_{\rm RH}) = \rho_\phi(T_{\rm RH})
\eeq
which gives, using Eqs.~(\ref{Eq:rhophi}) and (\ref{Eq:rhoRapp})
\beq
\frac{a_{\rm RH}}{a_{\rm end}}=\left[ \frac{k+8-6kl}{2k} \frac{M_P^{4l-1}\rho_{\rm end}^{\frac{1}{2}-l}}{\sqrt{3}\gamma_\phi}\right]^{\frac{k+2}{3k-6kl}} \, ,
\eeq
for $8+k-6kl >0$. For $8+k-6kl < 0$, we can use Eq.~(\ref{Eq:rhoRappter}) to obtain,
\beq
\frac{a_{\rm RH}}{a_{\rm end}}=\left[ \frac{6kl-k-8}{2k} \frac{M_P^{4l-1}\rho_{\rm end}^{\frac{1}{2}-l}}{\sqrt{3} \gamma_\phi}\right]^{\frac{k+2}{2k-8}} \, .
\label{arhaendneg}
\eeq
Note that Eq.~(\ref{arhaendneg}) is only true for $k>4$.
When $k \le 4$ and $8+k-6kl < 0$, reheating never occurs.
Consider for example the case for $\phi \phi \to bb$.
In Eq.~(\ref{arhaendneg}), we would find $a_{\rm RH} < a_{\rm end}$ which is clearly unphysical. Indeed, from Table~\ref{Tab:table}, for $k=2$ we infer that $\rho_R \sim a^{-4}$ while $\rho_\phi \sim a^{-3}$. For this case even for $k=4$, $\rho_R$ never comes to dominate the energy density in the absence of other inflaton-matter couplings.

For $k=2$, inflaton decays to fermions dominate at late times with respect to scatterings to bosons, and  $l=0$, so that
$\frac{a_{RH}}{a_{\rm end}} \sim 1.8 \times 10^{10}$ taking the parameter values used 
in Fig.~\ref{fig:mixedtempsk2}. 
Furthermore, for $y = 10^{-7}$, ${\cal R} < 1$ initially, and for $k=2$, it remains so, the reheat temperature is
not affected by the fermionic suppression. 
For $k=4$, boson final states dominate at late times, $l = -1/4$, and from the parameters used in Fig.~\ref{fig:mixedtemps1} we obtain
$\frac{a_{RH}}{a_{\rm end}}= 4 \times 10^5$. However, a more precise calculation should take into account the change in the slope of $\rho_R$
due to the kinematic enhancement when $\mathcal{R} > 1$. 
In this case, we obtain
\beq\label{eq:aRdef}
\rho_R = \frac{\mu_{\eff}^2 M_P \rho_{\rm end}^{\frac{1}{4}}}{36 \pi \lambda^{\frac{1}{4}}} \left( \frac{a_{\rm end}}{a_{\cal R}}\right)
\left( \frac{a_{\cal R}}{a}\right)^\frac{1}{2},
\eeq
where $a_{\cal R}$ is the scale factor from which the boosted enhancement
begins to have significant effect, computed in Eq.(\ref{Eq:k4r1}) (that is, $a= a_{\cal R}$ when ${\cal R} = 1$).  
Then the scale factor at reheating determined by $\rho_R=\rho_\phi$ is
\beq
\frac{a_{\rm RH}}{a_{\rm end}}= \left[\sqrt{\frac{a_{\cal R}}{a_{\rm end}}}
\frac{36 \pi \lambda^{\frac{1}{4}}\rho_{\rm end}^\frac{3}{4}}{\mu_{\eff}^2 M_P}\right]^{\frac{2}{7}} \sim 10^5,
\eeq
where we used Eq.~(\ref{Eq:k4r1}) for $\frac{a_{\cal R}}{a_{\rm end}}$. This result is in good agreement with Fig.~\ref{fig:mixedtemps1}.

When ${\cal R} < 1$, it is relatively straight forward to use the expressions for $a_{\rm RH}$ to determine
the reheating temperature:
\begin{align}
T_{\rm RH} \;&=\; \left(\frac{30}{g_\rho \pi^2} \right)^{\frac{1}{4}}
\left[\frac{2k}{k+8-6kl} 
\frac{\sqrt{3}\gamma_\phi}{M_P^{4l-1}} \right]^{\frac{1}{2-4l}}
\end{align}
for $8+k-6kl >0$. For $8+k-6kl < 0$ and $k> 4$,
\begin{align}
T_{\rm RH} \;&=\; \left(\frac{30}{g_\rho \pi^2} \right)^{\frac{1}{4}}
\left[\frac{2k}{6kl-k-8} 
\frac{\sqrt{3}\gamma_\phi}{M_P^{4l-1}} \rho_{\rm end}^{\frac{6kl-k-8}{6k}}\right]^{\frac{3k}{4k-16}} \, .
\end{align}
For the particular case depicted in Fig.~\ref{fig:mixedtempsk2},
for $\phi \rightarrow \bar f f$, $l = \frac12 - \frac1k$, and  we have
\beq
T_{\rm RH}^f=\left(\frac{30}{g_\rho \pi^2} \right)^{\frac{1}{4}}
\left[\frac{k \sqrt{3k(k-1)}}{7-k} \lambda^{\frac{1}{k}}
 \frac{y^2}{8 \pi}\right]^{\frac{k}{4}}M_P \, ,
\label{trf}
\eeq
and for the parameters used in Fig.~\ref{fig:mixedtempsk2} and
$g_\rho \sim 100$, 
$T_{\rm RH} \simeq 4.4 \times 10^7$ GeV.
For decays to bosons, $\phi \rightarrow bb$, $l = \frac1k-\frac12$, and we include the enhancement factor
proportional to ${\cal R}^{1/2}$ (which applies for $k>3$) and find
\beq
T_{\rm RH}^b=\left(\frac{30}{g_\rho \pi^2} \right)^{\frac{1}{4}}
 \left[\frac{1}{7 \pi^{3/2}} \sqrt{\frac{3}{k(k-1)^3}}\frac{\Gamma(\frac1k)}{\Gamma(\frac12+\frac1k)} \left(\frac{\mu}{\lambda^{1/k}M_P}\right)^{5/2} \left(\frac{\mu_{\eff}}{\mu}\right)^2 \right]^{\frac{k}{6k-10}}M_P\, .
\label{trb}
\eeq
When evaluated with the parameters used in Fig.~\ref{fig:mixedtemps1}, for $k=4$, we have $T_{\rm RH} = 2\times 10^{10}$ GeV.
Eqs.~(\ref{trf}) and (\ref{trb}) are two solutions for $T_{\rm RH}$ corresponding to cases considered in the examples in Figs.~(\ref{fig:mixedtempsk2}) and (\ref{fig:mixedtemps1}).
There are of course several other possible expressions for $T_{\rm RH}$ depending on the kinematic factor ${\cal R}$.
When ${\cal R} > 1$, we must modify the integrand used to determine Eq.~(\ref{Eq:rhoR}) as well as the limits of integration if ${\cal R}$ evolves in such a way that it crosses ${\cal R} = 1$ between $a_{\rm end}$ and $a_{\rm RH}$.

%%%%%%%%%%%%%%%%%%%%%%%%%%%
\section{Dark matter production}
\label{sec:dm}
%%%%%%%%%%%%%%%%%%%%%%%%%%%

As noted earlier, it is possible to produce certain very weakly interacting dark matter candidates during the reheating process. The relic abundance of these dark matter candidates may depend primarily on $T_{\rm max}$, $T_{\rm RH}$, or both
depending on the production cross section. 
We parametrize the thermally-averaged effective cross section for dark matter (DM) production in the following way,
\beq\label{eq:sigmav}
\langle \sigma v\rangle \;=\; \frac{T^{n}}{{\tilde \Lambda }^{n+2}}\,,
\eeq
where the mass scale ${\tilde \Lambda}$ is assumed to be parametrically related to the mass of a heavy mediator in the UV theory. For $n>-1$, DM production after reheating is subdominant~\cite{egnop,Garcia:2017tuj,GA,Elahi:2014fsa}.
For example, in the case of a weak scale gravitino, $n=0$, and ${\tilde \Lambda} \propto M_P$. In contrast, in high scale supersymmetry, 
$n=6$, and ${\tilde \Lambda}^2 \propto m_{3/2} M_P$. 
It is worth emphasizing that this effective description is valid as long as ${\tilde \Lambda}$ is above $T_{\rm max}$. The amount of DM produced during reheating is obtained from the solution of the following Boltzmann equation, 
\beq\label{eq:nchieom}
\dot{n}_{\chi} + 3Hn_{\chi} \;=\; g_{\chi}^2\langle \sigma v\rangle n_{r}^2 \;\equiv\; R(T) \,,
\eeq
where $g_{\chi}$ denotes the number of internal degrees of freedom of the DM particle $\chi$, and $n_R$ corresponds to the number density of the radiation, which in equilibrium can be written as
\beq
n_R \;=\; \frac{\zeta(3)}{\pi^2} T^3 \,.
\eeq
The production rate per unit volume can be written 
as 
\beq
R(T) = \frac{T^{n+6}}{\Lambda^{n+2}} \, ,
\eeq
where we have absorbed the numerical factors in 
$\Lambda^{n+2} = {\tilde \Lambda}^{n+2} \pi^4/g_\chi^2 \zeta(3)^2$.

Assuming instantaneous thermalization, it is convenient to define the DM yield as $Y_{\chi} \equiv n_{\chi}/T^{\frac{(4k+8)}{(k+2kl)}}$, where the power of $T$ is inferred from Eq.~(\ref{Eq:tfa}) with $Y_\chi \sim n_\chi a^3$.
The Boltzmann equation (\ref{eq:nchieom})
can be rewritten as
\beq
\frac{dY_\chi}{dT} = - \frac{R(T)}{H(T)} \left(\frac{4k+8}{3k+6kl} \right) T^{-\frac{5k+8+2kl}{k+2kl}} \,,
\label{dYdT}
\eeq
(if $8+k-6kl>0$). Furthermore, we can write $H(T)$ (which we assume is dominated by $\rho_\phi$) in terms of $T_{\rm RH}$
by noting that at $T_{\rm RH}$, $\rho_\phi = \rho_R$ and that 
$\rho_R(T_{\rm RH}) = \alpha T_{\rm RH}^4$, where $\alpha = g_\rho \pi^2/30$. Using the scaling of $\rho_\phi$ with $a$ from Eq.~(\ref{Eq:rhophi}), and the scaling of $a$ with $T$ 
from Eq.~(\ref{Eq:tfa}), we can write
\beq
H = \sqrt{\frac{\alpha}{3}} \frac{T_{\rm RH}^2}{M_P}
 \left(
\frac{T}{T_{\rm RH}} \right)^{\frac{4}{1+2l}} \, 
\eeq
which is interestingly independent of $k$ (except for the implicit $k$ dependence in $l$).

We are now in a position to integrate Eq.~(\ref{dYdT})
\beq
Y_{\chi}(T_{\rm RH}) = \sqrt{\frac{3}{\alpha}} \frac{M_P T_{\rm RH}^{\frac{2-4l}{1+2l}}}{\Lambda^{n+2}} \left(\frac{4k+8}{3k+6kl} \right) \int_{T_{\rm max}}^{T_{\rm RH}} dT \, T^{n+6} T^{-\frac{4}{1+2l}}  T^{-\frac{5k+8+2kl}{k+2kl}} \, ,
\label{eq:Ychiex}
\eeq
which is easily integrated to give
\beq
n_{\chi}(T_{\rm RH}) \;\simeq\; \sqrt{\dfrac{1}{3\alpha}} M_P \begin{cases}
\left( \dfrac{4k+8}{8+2k-12kl-kn-2kln} \right) \dfrac{T_{\rm RH}^{n+4}}{\Lambda^{n+2}} \,, & n < \dfrac{8+2k-12kl}{k(1+2l)} \,,\\[10pt]
\left( \dfrac{4k+8}{k+2kl} \right) \dfrac{T_{\rm RH}^{n+4}}{\Lambda^{n+2}} \ln\left(\dfrac{T_{\rm max}}{T_{\rm RH}}\right)\,, & n = \dfrac{8+2k-12kl}{k(1+2l)} \,, \\[10pt]
\left( \dfrac{4k+8}{12kl+kn+2kln-8-2k} \right)
\left( \dfrac{T_{\rm RH}}{T_{\rm max}} \right)^{\frac{8+6k-4kl}{k+2kl}} \dfrac{T_{\rm max}^{n+4}}{\Lambda^{n+2}}\,, & n > \dfrac{8+2k-12kl}{k(1+2l)} \, .
\end{cases}
\label{solY}
\eeq
For $l \to (k-2)/2k$ as in the first line of Eq.~(\ref{ls})
for fermionic final stats, these equations reduce to those in
\cite{Garcia:2020eof}. If we further specify $k=2$, they reduce to
the results in \cite{Garcia:2017tuj,Kaneta:2019zgw}. Note that aside from the prefactor, when $n \le \frac{8+2k-12kl}{k(1+2l)}$,
the abundance scales as $T_{\rm RH}^{n+4}$, ie., independent of $k$ and $l$. Only for larger $n$,
does the power of $\frac{T_{\rm RH}}{T_{\rm max}}$ depend on $k$ and $l$. Though one should bear in mind that both $T_{\rm RH}$ and $T_{\rm max}$ each depend on $k$ and $l$ as discussed in the previous section. 

Finally, the dark matter number density produced by scatterings in the thermal plasma
given in Eqs.~(\ref{solY}) can be converted to the dark matter contribution to the critical density using
\begin{align}
\Omega_{\chi}h^2 \;&=\; \frac{m_{\chi} n(T_0)}{\rho_c h^{-2}} \nonumber \\
&=\; \frac{\pi^2 g_{\rho}(T_0) m_{\chi} n_{\gamma}(T_0) n_\chi(T_{\rm RH})}{2\zeta(3)g_{\rho}(T_{\rm RH}) T_{\rm RH}^3 \rho_c h^{-2}} \nonumber \\
&=\; 5.9 \times 10^6 {\rm GeV}^{-1} \frac{m_{\rm DM} n_\chi(T_{\rm RH})}{T_{\rm RH}^3}
\label{oh2s}
\,,
\end{align}
where $g_{\rho}(T_0)=43/11$ is the present number of effective relativistic degrees of freedom for the entropy density, $n_{\gamma}(T_0)\simeq 410.66\,{\rm cm}^{-3}$ is the number density of CMB photons, and $\rho_c h^{-2}\simeq 1.0534\times 10^{-5}{\rm GeV\,cm}^{-3}$ is the critical density of the Universe. We consider for definiteness the high-temperature Standard Model value $g_{\rm RH}=427/4$.

Of course it is also possible that if the dark matter is coupled to the inflaton, that it may be produced directly in the decay process
\cite{egnop,Garcia:2017tuj,grav2,Garcia:2020eof}. However,
as we have seen, the decay to dark matter may be suppressed,
if the dark matter is fermionic, or enhanced if bosonic.
To fully treat the production of dark matter through decay,
we would need to specify separately a value of $l_\chi$ for the dark matter which may in principle differ from that of the 
standard model decay products involved in reheating. This is beyond the scope of the present work. Furthermore, even if the dark matter
is not directly coupled to the inflaton, 
but is coupled to the standard model, the production of dark matter
through direct decays may still proceed through loops \cite{Kaneta:2019zgw}, further complicating the calculation of the dark matter abundance. We save this study for future work.

\section{An example: the SUSY case}\label{sec:susy}

The general reheating formalism that we have developed in the previous sections can be applied to a wide variety of concrete models. In this section we implement it for a particular supersymmetric scenario. Consider the following form for the superpotential,
\beq\label{eq:Wsusy}
W \;=\; Y H_2 L \Phi + F(\Phi) + \cdots\,.
\eeq
Here $L=(\nu,\ell_L)$ denotes one of the three MSSM lepton doublets, $H_2$ is one of the two Higgs doublets, and SU(2) contractions are implicit. The inflaton superfield is denoted by $\Phi$, and $F(\Phi)$ represents the inflaton-sector interactions that lead to a potential of the form (\ref{eq:Vphi}). For example, one can assume a
superpotential\footnote{ We take $M_P=1$ in this expression.} of the form 
\beq
F  = 2^{\frac{k}{4}+1} \sqrt{\lambda} \left(\frac{\Phi^{\frac{k}{2}+1}}{k+2} 
-\frac{\Phi^{\frac{k}{2}+3}}{3(k+6)}
\right) \, ,
\eeq
with a K\"ahler potential of the no-scale form leads to the potential given in Eq.~(\ref{eq:Vphi}) \cite{Garcia:2020eof}. With the Yukawa coupling in Eq.~(\ref{eq:Wsusy}), the inflaton also plays the role of the right handed sneutrino.\footnote{For related models, see e.g.~\cite{eno9}.}
With the inflaton given by the real part of the scalar component of $\Phi$, $\phi=\sqrt{2}\,{\rm Re}\Phi$, the Lagrangian corresponding to (\ref{eq:Wsusy}) for $\phi\ll M_P$ takes the following form,
\begin{align}\notag
\mathcal{L} \;=\; &-\frac{Y}{\sqrt{2}}\phi\left( \bar{\tilde{H}}_2^+ \ell_L + \bar{\ell}_R\tilde{H}_2^+ - \bar{\tilde{H}}_2^0 \nu_L - \bar{\nu}_R \tilde{H}_2^0 \right) - Y \partial_{\Phi}F(\phi) \left( H_2^+ \tilde{\ell}_L - H_2^0 \tilde{\nu} + {\rm h.c.}\right)\\ \label{eq:LlHsusy}
& - \frac{1}{2}Y^2 \phi^2 \left( |\tilde{\ell}_L|^2 + |\tilde{\nu}|^2 + |H_2^+|^2 + |H_2^0|^2 \right)  - V(\phi) + \cdots\,.
\end{align}
Here it is worth recalling that, in the globally supersymmetric limit, $V(\phi)=|\partial_{\Phi}F(\phi)|^2$. The previous expression allows for the computation of the tree-level decay rate of the inflaton into fermions and scalars in a straightforward way, if one disregards the induced effective masses of the decay products. In order to take into account this kinematic effect, it is necessary to determine the corresponding mass eigenstates. We obtain
\beq\label{eq:meffsusy}
m_{\eff}^2 \;=\; \begin{cases}
\dfrac{1}{2}Y^2\phi^2\,,\quad & \text{fermions}\,,\\[10pt]
\dfrac{1}{2}Y^2\phi^2 \pm Y \partial_{\Phi}F(\phi)\,,\quad & \text{bosons}\,,
\end{cases}
\eeq
where the positive sign corresponds to the linear combinations of $(\tilde{\ell}_L,H_2^{+*})$ and $(H_2^0,\tilde{\nu}^*)$, and the negative sign to the orthogonal combinations of their complex conjugates. 
One can note that the second term in the bosonic mass is related to supersymmetry breaking, and in its absence, the masses of the fermionic and bosonic components are equal. For the decay of the inflaton into fermions, the total decay rate can be determined in a straightforward way for arbitrary $k$,
\beq
\sum_{f} \Gamma_{\phi\rightarrow \bar{f}f} \;=\; \frac{Y_{\eff}^2}{8\pi}m_{\phi}\,,
\eeq
where $Y_{\eff}$ is defined as in (\ref{eq:gammaff}), replacing $y\rightarrow Y$. For bosons, the presence of the $F$-dependent term in the effective mass makes the nature of the decay process dependent on the form of the inflaton potential.\par\bigskip

For a quadratic potential, that is, $\partial_{\Phi}F=\frac{1}{\sqrt{2}}m_{\phi}\phi$ near $\phi=0$, the three- and four-body processes $\phi\rightarrow bb^*$ and $\phi\phi\rightarrow bb^*$ occur, with rates
\begin{flalign} \label{eq:phitobsusy2}
& \text{($k=2$)} & \Cen{3}{
\begin{aligned}
\sum_b \Gamma_{\phi\rightarrow bb^*} \;&\simeq\; \frac{Y^2}{8\pi}m_{\phi}  \alpha_{\phi\rightarrow bb^*}(\mathcal{R}) \,,\\
\sum_b \Gamma_{\phi\phi\rightarrow bb^*} \;&\simeq\; \frac{Y^4 \rho_{\phi}}{16\pi m_{\phi}^3}  \alpha_{\phi\phi\rightarrow bb^*}(\mathcal{R})\,,
\end{aligned}}      &&  
\end{flalign} 
where the oscillation-averaged kinematic factors $\alpha_{\phi\rightarrow bb^*}$ and $\alpha_{\phi\phi\rightarrow bb^*}$ are defined in Appendix~\ref{sec:susykin}.  Fig.~\ref{fig:SusyAllK2A} shows the numerical solution for the instantaneous temperature during reheating in the case $k=2$ for a coupling $Y=10^{-7}$. In this case no kinematic suppression is present at any time, for any of the $\phi$-dissipation processes. The decay channels to fermions and to bosons have identical rates, which can be immediately appreciated in the figure. On the other hand, the scattering process $\phi\phi\rightarrow bb^*$ is always subdominant. Hence, the total decay rate is equal to twice the fermionic rate, and the temperature decreases during reheating as $T\propto a^{-3/8}$.

\begin{figure}[!ht]
\centering
    \includegraphics[width=0.65\textwidth]{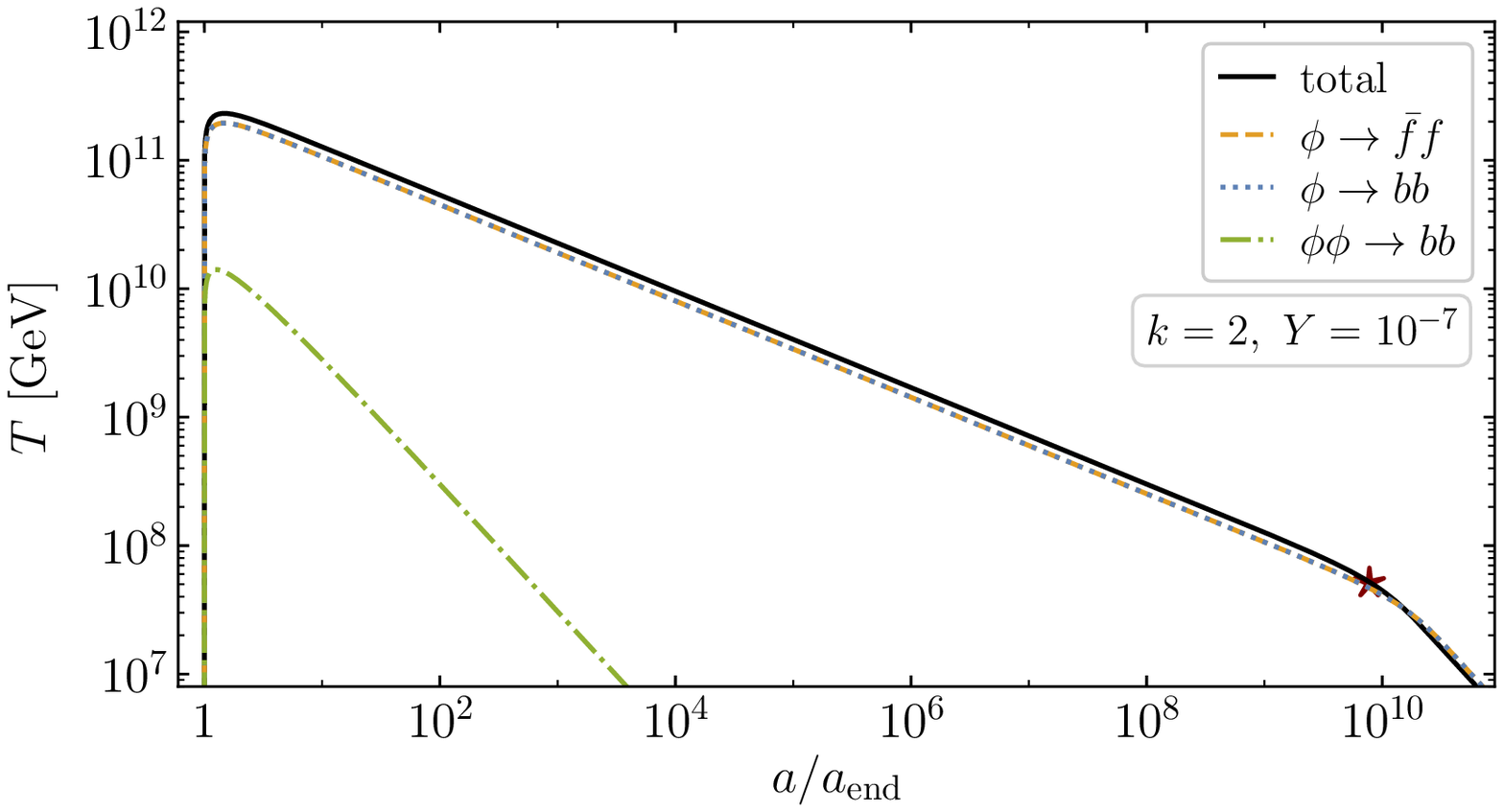}
    \caption{Instantaneous temperature as a function of the scale factor in the supersymmetric scenario (\ref{eq:LlHsusy}) with a quadratic inflaton potential. Here $m_{\eff}\neq0$ and $\lambda=2.5\times 10^{-11}$, assuming T-attractor inflation boundary conditions. The star signals inflation-radiation equality.}
    \label{fig:SusyAllK2A}
\end{figure}

\begin{figure}[!ht]
\centering
    \includegraphics[width=0.98\textwidth]{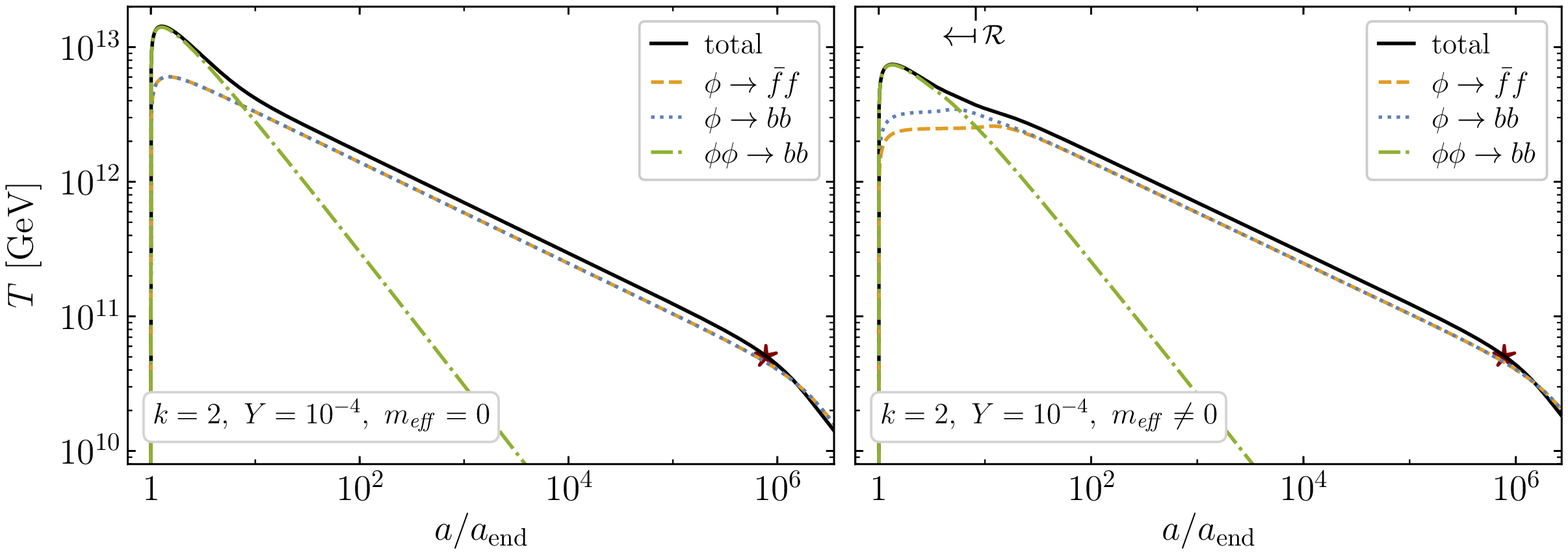}
    \caption{Instantaneous temperature as a function of the scale factor in the supersymmetric scenario (\ref{eq:LlHsusy}) with a quadratic inflaton potential. Left: $m_{\eff}=0$. Right: $m_{\eff}\neq0$. Here $\lambda= 2.3\times 10^{-11}$, assuming T-attractor inflation boundary conditions. The star signals inflation-radiation equality. The arrow points toward the region where $\mathcal{R}>1$ for one or more of the decay channels.}
    \label{fig:SusyAllK2B}
\end{figure}

Fig.~\ref{fig:SusyAllK2B} shows the scale factor dependence of the temperature for a larger value of the coupling\footnote{Note that at large couplings, our perturbative analysis begins to break down as discussed in the next section.}, $Y=10^{-4}$. The left panel depicts what the evolution of $T$ would be during reheating in the absence of oscillation-induced effective masses. We observe here the equality of fermion and boson decay rates, with a maximum temperature determined in this case by the scattering process, a scenario similar to that shown in Fig.~\ref{fig:mixedtempsk2}. The right panel in turn shows the resulting evolution $T(a)$ including the induced masses of the decay products. Recalling that for all three processes, for $k=2$, the suppression in the width decreases with time, we note that the differences with respect to the left panel are present only for $a\lesssim 20\,a_{\rm end}$. In this regime, the symmetry between the fermionic and bosonic rates is broken due to condensate effects, and $\phi\rightarrow bb^*$ dominates over $\phi\rightarrow \bar{f}f$. Nevertheless, despite the noticeable decrease in $T_{\rm max}$, by a factor of $\sim 1.9$, $\phi\phi\rightarrow bb^*$  controls the production of relativistic particles at very early times, as it does when $m_{\eff} = 0$.

For a quartic potential, with $\partial_{\Phi}F=\sqrt{\lambda}\phi^2$ near $\phi=0$, only the scattering process occurs at lowest order in the coupling $Y$, with a rate given by
\begin{flalign} \label{eq:phitobsusy4}
& \text{($k=4$)} & \Cen{3}{
\begin{aligned}
\sum_b \Gamma_{\phi\phi\rightarrow bb^*} \;=\;  \frac{Y^2\rho_{\phi}}{2.4\pi m_{\phi}^3} \Big\{& (Y+2\sqrt{\lambda})^2  \alpha_{\sigma}(\mathcal{R})_{\sigma\rightarrow Y(Y+ 2\sqrt{\lambda})/2} \\
& + (Y-2\sqrt{\lambda})^2  \alpha_{\sigma}(\mathcal{R})_{\sigma\rightarrow Y(Y - 2\sqrt{\lambda})/2} \Big\}\,,
\end{aligned}}      &&  
\end{flalign} 
(for details see Appendix~\ref{sec:susykin}). It is worth noting that, for the `minus' states with $Y\lesssim 2\sqrt{\lambda}$, an enhancement of the decay rate appears, instead of a suppression. Nevertheless, this enhancement is always $\lesssim 8\%$, and is tied to the process with the smallest branching ratio when it is maximized, making its contribution to $\Gamma_{\phi}$ negligible. The bosonic enhancement that increases with time is therefore not present in this supersymmetric construction.

\begin{figure}[!ht]
\centering
    \includegraphics[width=0.65\textwidth]{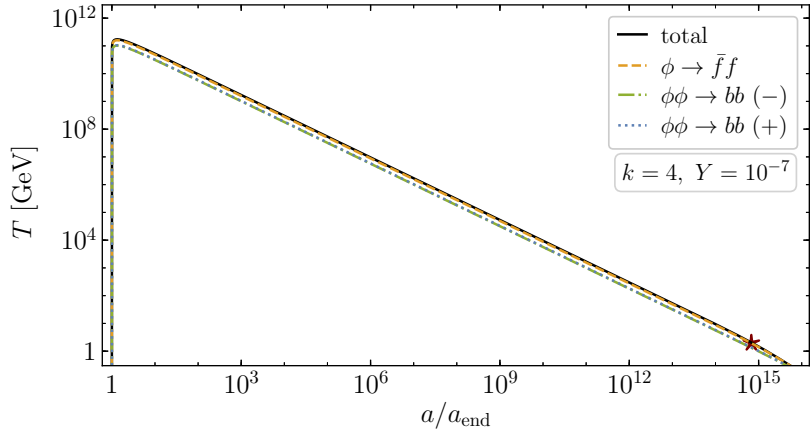}
    \caption{Instantaneous temperature as a function of the scale factor in the supersymmetric scenario (\ref{eq:LlHsusy}) with a quartic inflaton potential. Here $m_{\eff}\neq0$ and $\lambda= 3.3\times 10^{-12}$, assuming T-attractor inflation boundary conditions. The star signals inflation-radiation equality.}
    \label{fig:SusyAllK4A}
\end{figure}

Fig.~\ref{fig:SusyAllK4A} shows the temperature during reheating with $k=4$ for $Y=10^{-7}$. As we saw in  Fig.~\ref{fig:SusyAllK2A} for $k=2$, there is no appreciable enhancement or suppression in this case as well. Indeed, for all decay channels $\mathcal{R}\ll 1$. The inflaton decay rate for both effective bosonic channels is identical in this regime, and is suppressed by a factor of $\sim 5$ with respect to the fermionic one. Therefore, $T\propto a^{-3/4}$. A star, located at $a\simeq 7\times 10^{14} a_{\rm end}$, signals inflaton-radiation equality, that is, the end of reheating. Note the decrease of more than 7 orders of magnitude in $T_{\rm RH}$ compared to the quadratic case. 

\begin{figure}[!ht]
\centering
    \includegraphics[width=0.98\textwidth]{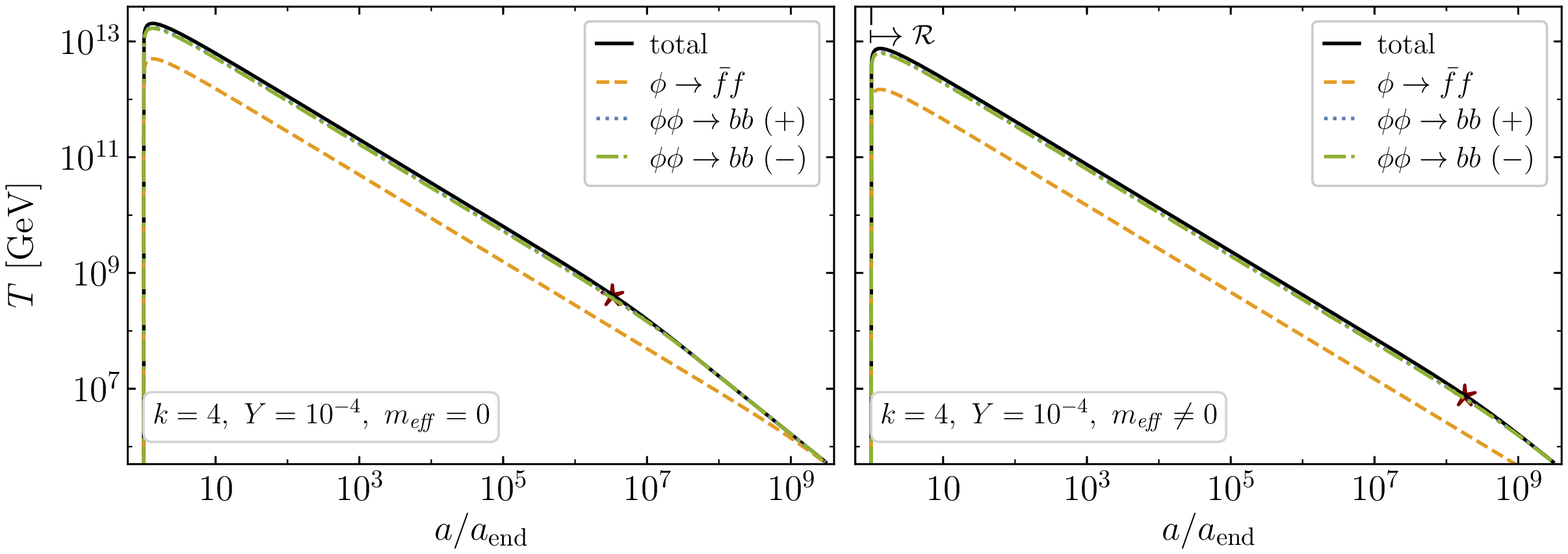}
    \caption{Instantaneous temperature as a function of the scale factor in the supersymmetric scenario (\ref{eq:LlHsusy}) with a quartic inflaton potential. Left: $m_{\eff}=0$. Right: $m_{\eff}\neq0$. Here $\lambda= 3.3\times 10^{-12}$, assuming T-attractor inflation boundary conditions. The star signals inflation-radiation equality. The arrow points toward the region where $\mathcal{R}>1$ for one or more of the decay channels.}
    \label{fig:SusyAllK4C}
\end{figure}

The choice of $Y=10^{-4}$ with $k=4$ is displayed in Fig.~\ref{fig:SusyAllK4C}. For this coupling, all decay channels acquire a kinematic suppression. Analogously to the $k=2$ case, the left panel shows the resulting temperature disregarding the inflaton-induced masses for the fermionic and bosonic decay products. Unlike the previous cases though, here it is the scattering of $\phi$ into bosons that most efficiently heats the Universe, to a reheating temperature $T_{\rm RH}\simeq 4\times 10^8\,{\rm GeV}$. In the right panel we observe the effect of the kinematic suppression. The fermionic width is reduced by a factor of $7\times 10^{-3}$, while the dominant bosonic widths acquire a suppression $\simeq 2\times 10^{-2}$. This reduction is time-independent, and results in $T_{\rm RH}\simeq 8\times 10^6\,{\rm GeV}$.

\section{Discussion}
\label{sec:disc}

\subsection{Limitations}\label{sec:limitations}

Before our summary, we point out a few of the limitations
of our analysis. Throughout, we have assumed perturbative reheating processes. 
Furthermore, we have assumed that the mass of the decay products are genuinely given by the inflaton condensate during the reheating. Thus during an oscillation period, the mass necessarily passes through zero, which may trigger non-perturbative particle production.
Whether the non-perturbative production becomes dominant source for $\rho_R$ or not depends on the strength of the couplings.
For instance, in the case of $\phi\to \bar f f$,  non-perturbative production becomes non-negligible for $y\gtrsim 10^{-6}$ when $k=2$, and for $y\gtrsim 0.1\sqrt{\lambda}$ when $k=4$~\cite{Greene:1998nh}.
These values happen to be close to those obtained from ${\cal R}>1$, as can we discuss further in the Appendix. Similar limits for $\phi\to bb$ and $\phi\phi\to bb$ apply and can also be approximated from ${\cal R}>1$ (see, for instance, \cite{Dufaux:2006ee}).

Also, we have not incorporated the thermal corrections to the decay products, which may cause a similar kinetic suppression when the temperature is sufficiently high, compared to the inflaton mass.
Although for $k=2$, $m_\phi$ is constant, for $k>2$ $m_\phi$ decreases slower than the temperature, with $m_\phi\propto a^{-1}~(k=4)$ and $a^{-3/2}~(k=6)$.
Therefore, for $\phi\to \bar f f$ with $k=4$, we obtain
\bea
\frac{m_{\rm th}^2}{m_\phi^2}\sim \frac{g^2y_{\rm eff}}{2(\lambda^3\rho_{\rm end})^{1/8}}\sqrt{\frac{5M_P}{6g_\rho\pi^3}}\left(\frac{a}{a_{\rm end}}\right)^{3/2} \gtrsim 1
\quad\ \Rightarrow \quad\ 
a\gtrsim 1200\,a_{\rm end}\,,
\eea
for the parameter choices used in Fig.~\ref{fig:mixedtemps1}, and the adopted thermal mass is $m_{\rm th}\sim g T$ with $g\sim 0.1$. 
When one considers the cases where thermal masses are greater than the inflaton mass, the thermal dissipation rate of the inflaton should carefully be taken into account, since the inflaton energy density can still be transferred into radiation \cite{Yokoyama:2005dv}.

For $k\neq 2$, the non-perturbative excitation of non-vanishing momentum modes of the inflaton can also be sourced by the self-interaction of $\phi$. Not only this can have a significant effect on the production rate of daughter particles, but it can also lead to the fragmentation of the inflaton condensate. It has been shown \cite{Lozanov:2016hid,Lozanov:2017hjm} that neglecting inflaton-matter couplings for a potential of the form (\ref{eq:attractor}), disrupts the condensate occurs through a narrow self-resonance, and leads to its fragmentation after $\mathcal{O}(5)$ $e$-folds for $k=4$. Similar conclusions can be derived in the presence of the four-body interaction $\phi^2b^2$~\cite{Maity:2018qhi}. Nevertheless, estimating the duration of the condensate in the presence of generic matter decay channels, and its effect on the dark matter abundance, lies  beyond the scope of this work.

\subsection{Summary}

An essential feature of any inflationary model
is its ability to reheat the Universe after a period of exponential expansion.  In some cases, it is sufficient to know that the Universe reheat to 
a certain temperature $T_{\rm RH}$ and that equilibrium was established, and a period of radiation dominated expansion ensued. If the 
reheat temperature is sufficiently high to allow for baryogenesis, nucleosynthesis, and the production of weakly interacting dark matter the details of the reheating process may not be important. However, for the production of superweakly interacting dark matter, which never attains thermal equilibrium, these details may
be essential for determining the relic abundance. 

When one goes beyond the instantaneous reheating approximation, one finds that if thermalization is sufficiently rapid, the first few decay products begin to heat the Universe to a temperature, $T_{\rm max}$, which could be much larger than $T_{\rm RH}$ \cite{Giudice:2000ex,Garcia:2017tuj},
though the energy density in the newly created radiation bath is much less than that stored in the ongoing inflaton oscillations. Typically we expect
that during reheating the temperature falls off slowly with the expansion, $T \sim a^{-3/8}$ as
more energy is pumped into the thermal bath from
continual decays. Dark matter production may be sensitive to the maximum temperature (and the evolution to down to $T_{\rm RH}$) if its production cross section is proportional to $T^n$ with $n>6$.

In this paper, we have considered several important aspects of the reheating process. First, 
the evolution of the temperature, $T(a)$ depends
on the form of the inflaton potential controlling the period of inflaton oscillations \cite{Bernal:2019mhf,Garcia:2020eof}. For $V(\phi) \propto \phi^k$, the evolution is certainly sensitive to $k$ which affects the equation of state during oscillations. However, for $k>2$, 
the mass of the inflaton and hence its decay rate
are also dependent on $k$ and also affects the evolution of the temperature.  Second, the evolution depends on the spin statistics of the dominant
final state of inflaton decay. We have shown
that the temperature evolution differs depending 
on whether the inflaton decays predominantly to fermions or bosons, or whether annihilations to boson are the main channel depleting the oscillations. For $k=2$, $T \sim a^{-3/8}$ for both
fermionic and bosonic final states, and annihilations are incapable of reheating the Universe. However, for $k>2$, the behaviour differs, and annihilations are capable of reheating. Third, while we noted that for $k>2$,
the inflaton mass also undergoes a damped oscillation, for all $k$, the masses of the final state particles also depend on the inflaton field value. This may cause the effective final state mass to exceed the inflaton mass and lead to 
a suppression in the decay rate (for fermionic final states and annihilations) or an enhancement (for bosonic final states). We have used the thermal production of dark matter as an application of these results. 

There are several stones left unturned in our analysis. For sufficiently large couplings,
non-perturbative effects can not be neglected.
Thus parametric resonance may also play a role in the reheating process. We have also set aside the question of direct couplings of the inflaton to dark matter. In this case, the evolution and abundance of dark matter will depend on both the statistics of final state initializing the thermal bath, and the spin of the dark matter particle. 
The thermal contribution to final state masses
may also be important. Finally, it is important to 
re-examine the validity of the instantaneous thermalization approximation used in this work.

\section*{Acknowledgements}

This work was made possible by Institut Pascal at Universit\'e
Paris-Saclay with the support of the P2I and SPU research departments and 
the P2IO Laboratory of Excellence (program ``Investissements d'avenir''
ANR-11-IDEX-0003-01 Paris-Saclay and ANR-10-LABX-0038), as well as the
IPhT. The work of MG was supported by the Spanish Agencia Estatal de Investigación through Grants No.~FPA2015-65929-P (MINECO/FEDER, UE) and No.~PGC2018095161-B-I00, IFT Centro de Excelencia Severo Ochoa SEV-2016-0597, and Red Consolider MultiDark FPA2017-90566-REDC. The work of K.A.O.~was supported in part by DOE grant DE-SC0011842  at the University of
Minnesota. The work of KK was supported by a KIAS Individual Grant (Grant No. PG080301) at Korea Institute for Advanced Study. MG would like to thank CNRS and the Laboratoire de Physique des 2 Infinis Irène Joliot-Curie for their hospitality and financial support
of the IN2P3 master project "Hot Universe and Dark Matter" while completing this work.

\appendix

\section{The Boltzmann equation for a decaying condensate}\label{sec:alt}

In this appendix we derive the evolution equation for the energy density of the inflaton condensate from the particle perspective. Under the assumption that the decay of the inflaton is perturbative, then the condensate, $\phi$, is spatially homogeneous, and its phase space distribution may be written as $f_{\phi}(k,t)=(2\pi)^3n_{\phi}(t)\delta^{(3)}(\boldsymbol{k})$, with $n_{\phi}$ the instantaneous inflaton number density. Disregarding Bose enhancement / Pauli blocking effects for the $\phi$-decay products, the integrated Boltzmann equation for the number density can be written as follows~\cite{Nurmi:2015ema},
\beq
\dot{n}_{\phi} + 3Hn_{\phi} \;=\; -\int d\Psi_{\phi,A,B}\,|\mathcal{M}|^2_{\phi\rightarrow AB} f_{\phi}(k,t)\,,
\eeq
where $A,B$ denote the decay products of $\phi$, $d\Psi_{\phi,A,B}$ is the phase space measure for the particles $A,B$ and the condensate $\phi$, and $\mathcal{M}$ denotes the transition amplitude.\footnote{Note that there is no back-reaction
producing inflatons in the condensate.
The effect of producing inflaton particle states from the back-reaction, as well as from the direct decay of the inflaton can also be neglected.} More precisely,
\beq
d\Psi_{\phi,A,B }\,|\mathcal{M}|^2_{\phi\rightarrow AB} \;=\; \sum_{n=1}^{\infty} \frac{d^3 \bk}{(2\pi)^3 n_{\phi}(t)} \frac{d^3\bp_A}{(2\pi)^32p_A^0} \frac{d^3\bp_B}{(2\pi)^32p_B^0} (2\pi)^4 \delta^{(4)}(p_n - p_A - p_B) |\mathcal{M}_n|^2\,,
\eeq
where now $\mathcal{M}_n$ denote the transition amplitude in one oscillation for each oscillating field mode of $\phi$ from the coherent state $|\phi\rangle$ to the two-particle final state $|A,B\rangle$. Below we perform a few explicit computations for it. Note here that the seemingly Lorentz non-invariant factor of $1/n_\phi$ is needed so that the inflaton measure is correctly normalized by
\beq
\int \frac{d^3\bk}{(2\pi)^3 n_{\phi}}\,f_{\phi}(k,t) \;=\;1\,.
\eeq
Integration with respect to the $\phi$ momentum gives
\beq
\dot{n}_{\phi} + 3Hn_{\phi} \;=\; - \sum_{n=1}^{\infty} \int  \frac{d^3\bp_A}{(2\pi)^32p_A^0} \frac{d^3\bp_B}{(2\pi)^32p_B^0} (2\pi)^4 \delta^{(4)}(p_n - p_A - p_B) |\mathcal{M}_n|^2\,,
\label{matrix}
\eeq
where now $p_n=(E_n,\boldsymbol{0})$, $E_n$ denoting the energy of the $n$-th oscillation mode of $\phi$.
Note that the matrix element in Eq.~(\ref{matrix}) effectively contains the condensate, thereby absorbing the factor of $n_\phi$ that would be expected to be present on the right-hand side of (\ref{matrix}).

The evolution equation for the energy density of $\phi$ can be obtained by noting that, on the right-hand side, we must introduce the energy of each oscillation mode $|\mathcal{M}_n|^2\rightarrow |\mathcal{M}_n|^2 E_n$. On the left side of the equality, the adiabaticity assumption for the decay (cf.~(\ref{Eq:rhophi2})) implies that only the lowest oscillation mode must be taken into account, so that $\rho_{\phi}=m_{\phi} n_{\phi}$, where $m_{\phi}$ has been defined in (\ref{eq:mephiff}). Developing the time derivative we find 
\beq\label{eq:dotrhophi}
\dot{\rho}_{\phi} \;=\; -3H\left(-\frac{\partial_t m_{\phi}^2}{6Hm_{\phi}^2}\right)\rho_{\phi} + m_{\phi}\dot{n}_{\phi}\,.
\eeq
To simplify the first term on the right-hand side of the equality, we note that, from Eqs.~(\ref{Eq:rhophi2}) and (\ref{eq:rhoeom1}), it is straightforward to verify that the equation of motion for $\phi_0$ on short time-scales is given by
\beq
\dot{\phi}_0 \;\simeq\; -\frac{6}{k+2}H\phi_0\,.
\eeq
From the definition of the effective mass, Eq.~(\ref{eq:mephiff}), we then have
\beq
-\frac{\partial_t m_{\phi}^2}{6Hm_{\phi}^2} \;=\; \frac{(2-k)\dot{\phi}_0}{6H\phi_0} \;=\; \frac{k-2}{k+2}\,.
\eeq
Therefore, upon comparison with (\ref{eq:wk}), and substitution of (\ref{matrix}), we obtain
\beq
\dot{\rho}_{\phi} + 3H(1+w_{\phi})\rho_{\phi} \;=\; -(1+w_{\phi})\Gamma_{\phi}\rho_{\phi}\,, 
\eeq
where the right-hand side is given by the energy transfer per space-time volume (Vol$_4$), defined as
\bea
&&
(1+w_\phi)\Gamma_\phi\rho_\phi \equiv \frac{\Delta E}{{\rm Vol}_4},
\\
&&
\Delta E \equiv \int\frac{d^3\bp_A}{(2\pi)^32p_A^0} \frac{d^3\bp_B}{(2\pi)^32p_B^0}(p_A^0+p_B^0)|\langle {\rm f}|i\int d^4x{\cal L}_I|0\rangle|^2
\label{eq:DeltaE}
\eea
with ${\cal L}_I$ being the interaction Lagrangian.\footnote{We denote the initial state by $|0\rangle$, since there are no inflaton quanta produced when $t\to-\infty$.
Instead, $\phi$'s in ${\cal L}_I$ are treated as a time-dependent coefficient of the interaction, namely, $\phi$ being regarded as the homogeneously oscillating classical field.
Therefore, in computing the matrix element, we do not have symmetry factors arising from the initial state, as it is assumed to be vacuum. }
Substituting
\bea
|\langle {\rm f}|i\int d^4x{\cal L}_I|0\rangle|^2 = {\rm Vol}_4\sum_{n=-\infty}^\infty|{\cal M}_n|^2(2\pi)^4\delta^4(p_n-p_A-p_B),
\eea
to Eq.~(\ref{eq:DeltaE}), we obtain~\cite{Ichikawa:2008ne,Kainulainen:2016vzv}
\bea
&&
\Gamma_\phi=\frac{1}{8\pi(1+w_\phi)\rho_\phi}\sum_{n=1}^\infty|{\cal M}_n|^2E_n\beta_n(m_A,m_B),
\label{eq:gammageneral}
\\
&&
\beta_n(m_A,m_B)\equiv
\sqrt{\left(1-\frac{(m_A+m_B)^2}{E_n^2}\right) \left(1-\frac{(m_A-m_B)^2}{E_n^2}\right)}.
\eea

We may write $E_n=n\omega$, with $\omega$ being the frequency of oscillation of $\phi$, which decreases with the envelope $\phi_0$ for $k>2$. With $\phi(t) \simeq \phi_0(t)\cdot \mathcal{P}(t)$, approximating $\phi_0$ as constant over one oscillation we obtain
\beq
\dot{\mathcal{P}}^2 \;=\; \frac{2\rho_{\phi}}{\phi_0^2}\left(1-\mathcal{P}^k\right) \;=\; \frac{2m_{\phi}^2}{k(k-1)}\left(1-\mathcal{P}^k\right)\,.
\eeq
Straightforward integration gives
\beq
\omega \;=\; m_{\phi}\sqrt{\frac{\pi k}{2(k-1)}}\,\frac{\Gamma(\frac{1}{2}+\frac{1}{k})}{\Gamma(\frac{1}{k})}\,.
\eeq
In term of this frequency we can write
\beq
\mathcal{P}(t) \;=\; \sum_{n=-\infty}^{\infty} \mathcal{P}_n e^{-in \omega t}.
\eeq

\subsection{Inflaton decay to a pair of fermions}

Let us evaluate explicitly the decay rate for the energy density of $\phi$ when it decays to a pair of fermions. 
Assume now an inflaton-matter coupling of the form $\mathcal{L}_I=y\phi\bar{f}f$. 
In the amplitude ${\cal M}_n$ we replace $\phi$ with $\phi_0 {\cal P}_n$ (as it is treated as an interaction coefficient) and obtain
\bea
{\cal M}_n = y\phi_0 {\cal P}_n \bar u(p_A) v(p_B),
\eea
and thus, averaging over oscillations,
\begin{align} \notag \displaybreak[0]
\Gamma_{\phi\rightarrow \bar{f}f} \;&=\; \frac{y^2 \phi_0^2 \omega^3}{4\pi(1+w_{\phi})\rho_{\phi}} \sum_{n=1}^{\infty} n^3 |\mathcal{P}_n|^2 \left\langle \beta_n^{3}(m_f,m_f)_+ \right\rangle\\ \notag \displaybreak[0]
&=\; \frac{y^2 }{8\pi} \omega \left[ (k+2)(k-1) \left(\frac{\omega}{m_{\phi}}\right)^2 \sum_{n=1}^{\infty} n^3 |\mathcal{P}_n|^2 \left\langle \left(1 - \left(\frac{2 m_f}{n\omega}\right)^2 \right)^{3/2}_+ \right\rangle \right]\\\notag \displaybreak[0]
&=\; \frac{y^2 }{8\pi} \omega \left[ (k+2)(k-1) \left(\frac{\omega}{m_{\phi}}\right)^2 \sum_{n=1}^{\infty} n^3 |\mathcal{P}_n|^2 \left\langle \left(1 - \frac{\mathcal{R}}{n^2}\mathcal{P}^2 \right)^{3/2}_+ \right\rangle \right]\\ \label{eq:gammaphitofffull}
&\equiv\; \frac{y^2 }{8\pi} \omega\, \alpha_y(k,\mathcal{R})\,.
\end{align} 
Note here that $\mathcal{R}=(2m_f/\omega)^2|_{\phi\rightarrow \phi_0}$ (see Eq.~(\ref{eq:kincond})). Here we have introduced the notation
\beq
(1-x)_+ \;\equiv\; (1-x)\,\theta(1-x)\,,
\eeq
where $\theta$ is the Heaviside step function. This ensures that the decay only occurs when it is (instantaneously) kinematically allowed. Comparing (\ref{eq:gammaphitofffull}) and (\ref{eq:gammaff}), we finally identify
\beq\label{eq:yeffdef}
y_{\eff}^2(k) \;=\; \alpha_y(k,\mathcal{R})\frac{\omega}{m_{\phi}} y^2\,.
\eeq

\begin{figure}[!ht]
\centering
    \includegraphics[width=0.6\textwidth]{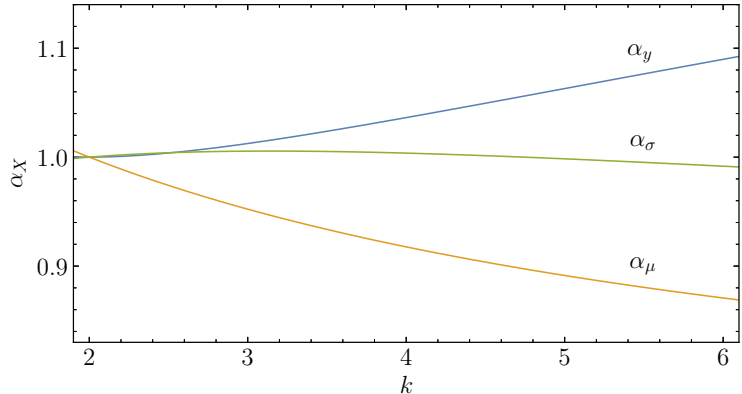}
    \caption{Numerical value of the sub-leading corrections to the effective inflaton matter-couplings $y$, $\mu$ and $\sigma$. These functions are computed as sums of coefficients of the Fourier expansion of powers of the exact solution of the equation of motion for $\phi$. Here $m_{\eff}=0$.}
    \label{fig:alphas}
\end{figure}

\begin{figure}[!ht]
\centering
    \includegraphics[width=0.65\textwidth]{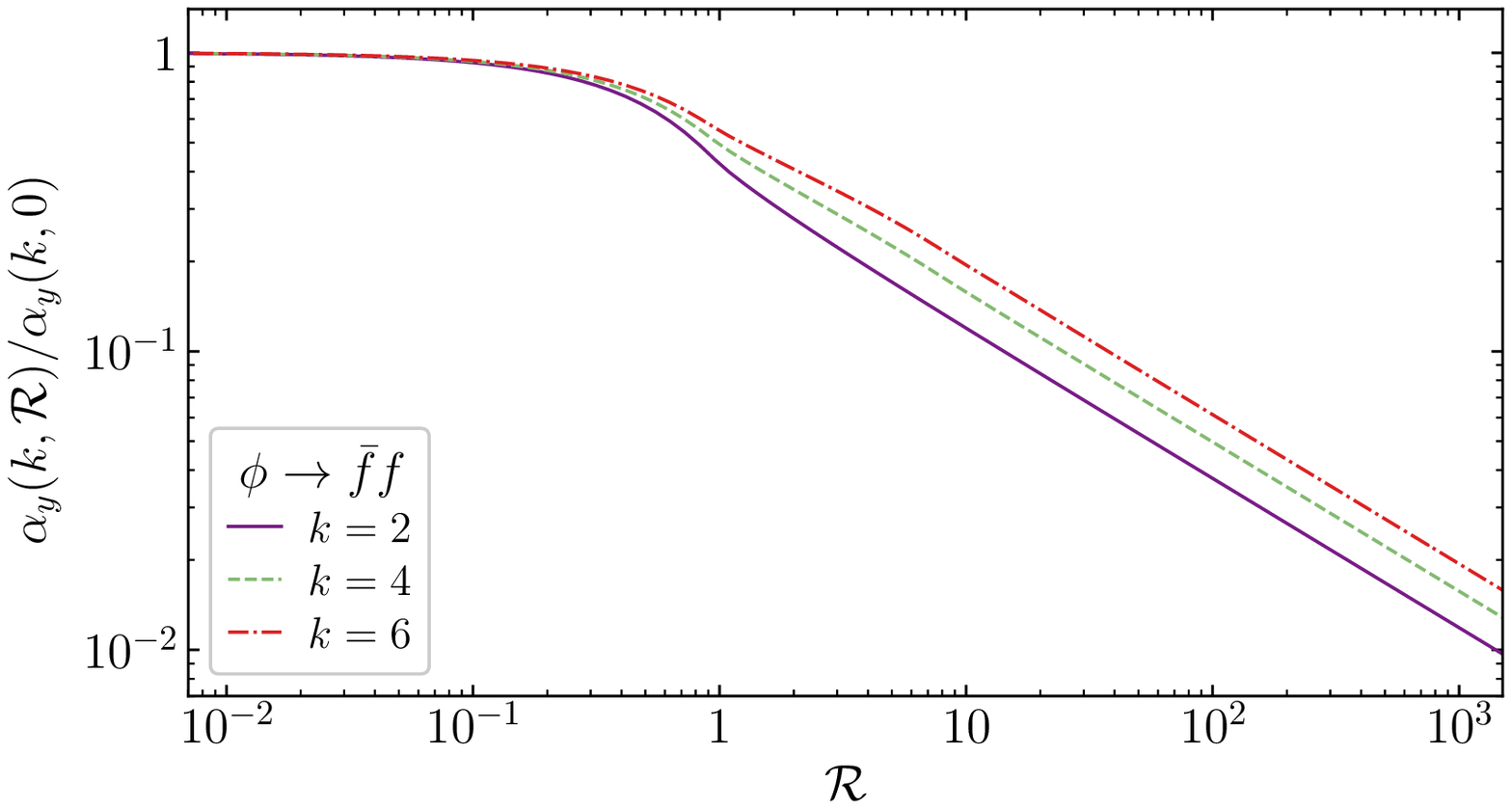}
    \caption{Kinematic factor for the oscillation-average of the decay rate $\Gamma_{\phi\rightarrow\bar{f}f}$. }
    \label{fig:KinF}
\end{figure}

Fig.~\ref{fig:alphas} shows the $k$-dependence of the function $\alpha_y(k,0)$, that is, in the massless fermion limit. Realistically though, $m_f\neq 0$ even if in the vacuum $m_f=0$. The dependence on the effective mass of $f$, induced by the oscillating background, is quantified through $\mathcal{R}$ and shown in Fig.~\ref{fig:KinF} for $k=2,4,6$.
For $\mathcal{R}\ll 1$, the rate at $m_f=0$ can be used. For $\mathcal{R}\gtrsim \mathcal{O}(10^{-1})$ the inflaton-induced mass for $f$ becomes comparable to $m_{\phi}$, and for $\mathcal{R}\gg 1$  
the effective decay rate is suppressed as $\alpha_y(k,\mathcal{R}) \propto \mathcal{R}^{-1/2} \alpha_y(k,0)$ (or $\Gamma_{\phi\rightarrow \bar{f}f} \propto \mathcal{R}^{-1/2}$). The constant of proportionality is approximately $0.38$ for $k=2$, $0.50$ for $k=4$ and $0.61$ for $k=6$. It is worth noting that, although the kinematic suppression is less severe for higher harmonics in (\ref{eq:gammaphitofffull}), this relative enhancement of the rate is over-compensated by the exponential suppression of the $\mathcal{P}_n$. In fact, approximating the kinematic blocking by means of the first harmonic only, a maximum error of $\sim 20\%$ is made (for $k=6$ and $\mathcal{R}\gg 1$).

\subsection{Inflaton decay to a pair of bosons}\label{app:phitobb}

Let us now consider an inflaton-matter coupling of the form $\mathcal{L}_I = \zeta^{2-\gamma} \phi^{\gamma} b_1 b_2$, where $\gamma$ is an arbitrary integer exponent. 
Again, we may replace $\phi^\gamma$ with $\phi_0^\gamma ({\cal P^\gamma})_n$ in ${\cal M}_n$, where the expression $\left(\mathcal{P}^{\gamma}\right)_n$ denote the Fourier coefficients of the expansion of $\mathcal{P}^{\gamma}(t)$, and obtain $\mathcal{M}_n= \zeta^{2-\gamma}\phi_0^{\gamma} \left(\mathcal{P}^{\gamma}\right)_n$. Substitution of the squared amplitude in (\ref{eq:gammageneral}) immediately gives
\beq
\Gamma_{\phi^{\gamma}\rightarrow b_1 b_2} \;=\; \frac{\zeta^{4-2\gamma}\phi_0^{2\gamma}}{8\pi(1+w_{\phi})\rho_{\phi}}\omega \sum_{n=1}^{\infty} n| \left(\mathcal{P}^{\gamma}\right)_n |^2 \left\langle \beta_n(m_1,m_2)_+ \right\rangle\,.
\eeq

\begin{figure}[!ht]
\centering
    \includegraphics[width=0.65\textwidth]{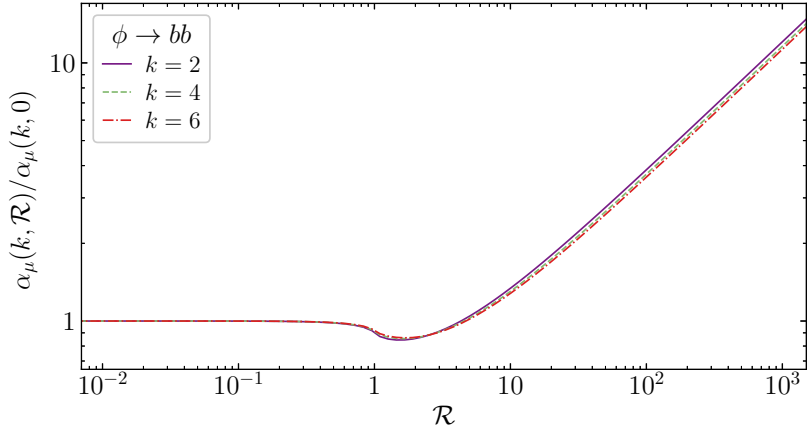}
    \caption{Kinematic factor for the oscillation-average of the decay rate $\Gamma_{\phi\rightarrow bb}$.}
    \label{fig:KinB1}
\end{figure}

For the case $\phi\rightarrow bb$, with $\gamma=1$ and $\zeta=\mu$, and including the appropriate symmetry factors, we obtain (\ref{eq:gammapbb}), with 
\begin{align} \notag
\mu_{\eff}^2(k) \;&=\; \frac{1}{4}(k+2)(k-1)\frac{\omega}{m_{\phi}} \left[ 4 \sum_{n=1}^{\infty} n| \mathcal{P}_n |^2 \left\langle \left(1-\frac{\mathcal{R}}{n^2}\mathcal{P} \right)^{1/2}_+ \right\rangle \right] \mu^2\\ \label{eq:zeffdef}
&\equiv\; \frac{1}{4}(k+2)(k-1)\frac{\omega}{m_{\phi}} \alpha_{\mu}(k,\mathcal{R}) \mu^2\,.
\end{align} 
Here $\mathcal{R}=(2m_b/\omega)^2|_{\phi\rightarrow\phi_0}$ is also given by the middle line of (\ref{eq:kincond}). The numerically computed function $\alpha_{\mu}$ in the limit $m_{b}\ll m_{\phi}$ is shown in Fig.~\ref{fig:alphas}. One can account for the inflaton-induced effective mass of $b$, assuming for simplicity a vanishing bare mass. The result of the numerical calculation of the average over one oscillation is shown in Fig.~\ref{fig:KinB1}. Notably, in this case, the approximation which uses the first harmonic accounts for $\gtrsim 99\%$ of the kinematic effect. As expected, for low values of the inflaton-matter coupling, $\mathcal{R}\ll 1$, the induced mass $m_{\eff}$ can be safely neglected. However, as $\mathcal{R}\gtrsim 1$, the value of $\alpha_{\mu}(k,\mathcal{R})$ deviates from that of $\alpha_{\mu}(k,0)$, and in fact appears to grow as $\mathcal{R}^{1/2}$ for $\mathcal{R}\gtrsim 5$. Numerically, this occurs due to the linear dependence on $\phi$ of the effective mass of $b$ (c.f.~Eq.~(\ref{eq:mprod})), meaning that for sufficiently large $\mathcal{R}$ the argument of the square root in (\ref{eq:zeffdef}) will be positive for half of the oscillation. Physically, this signals the breakdown of perturbativity and the need to account for (tachyonic) preheating effects.  \par\medskip

\begin{figure}[!ht]
\centering
    \includegraphics[width=0.65\textwidth]{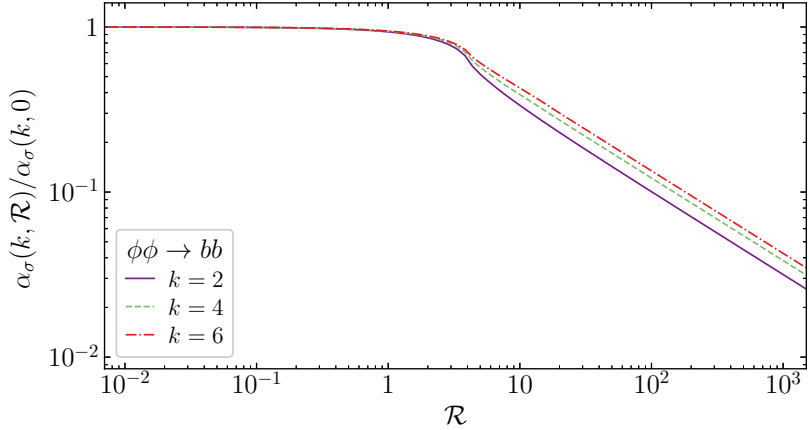}
    \caption{Kinematic factor for the oscillation-average of the decay rate $\Gamma_{\phi\phi\rightarrow bb}$.}
    \label{fig:KinB2}
\end{figure}

In the ``scattering'' scenario $\phi\phi\rightarrow bb$, using ${\cal L}_I=\sigma \phi^2b^2$, we obtain ${\cal M}_n=2\sigma\phi_0^2({\cal P}^2)_n$, and thus 
\bea
\Gamma_{\phi\phi\to bb}=\frac{\sigma^2\phi_0^4}{4\pi(1+w_\phi)\rho_\phi}\omega\sum_{n=1}^{\infty}n|({\cal P}^2)_n|^2 \left\langle \beta_n(m_b,m_b)_+ \right\rangle.
%\nonumber
\eea
For $k=2$, using $\sum_{n=1}^{\infty}n|({\cal P}^2)_n|^2=1/8$ and $\phi_0^4=(2\rho_\phi/m_\phi^2)^2$, we recover $\Gamma_{\phi\phi\to bb}=\sigma^2\rho_\phi/8\pi m_\phi^3$ when $m_b=0$.
We analogously obtain that 
\begin{align} \notag
\sigma^2_{\eff}(k) \;&=\; \frac{1}{8} k(k+2)(k-1)^2 \frac{\omega}{m_{\phi}} \left[ 8 \sum_{n=1}^{\infty} n| \left(\mathcal{P}^2\right)_n |^2 \left\langle \left( 1-\frac{\mathcal{R}}{n^2}\mathcal{P}^2 \right)^{1/2}_+ \right\rangle  \right] \sigma^2\\ \label{eq:seffdef}
&\equiv\; \frac{1}{8} k(k+2)(k-1)^2 \frac{\omega}{m_{\phi}} \alpha_{\sigma}(k,\mathcal{R}) \sigma^2\,,
\end{align} 
where the numerically calculated $\alpha_{\sigma}(k,0)$ is shown in Fig.~\ref{fig:alphas}. Here $\mathcal{R}=(2m_b/\omega)^2|_{\phi\rightarrow\phi_0}$, also given in Eq.~(\ref{eq:kincond}). Similarly to the fermionic decay case, the presence of the kinematic factor will result in a suppression of the effective decay rate of $\phi$ at large $\mathcal{R}$. Fig.~\ref{fig:KinB2} shows this effect for $k=2,4,6$. For $\mathcal{R}\ll 1$, the limit $m_b=0$ is appropriate. For $\mathcal{R}\gg1$, $\alpha_{\sigma}(k,\mathcal{R}) \propto \mathcal{R}^{-1/2} \alpha_{\sigma}(k,0)$, with constant of proportionality equal to $1.00$, $1.22$ and $1.35$ for $k=2,4$ and $6$, respectively.

\subsection{Supersymmetric kinematic factors}\label{sec:susykin}

\begin{figure}[!ht]
\centering
    \includegraphics[width=0.65\textwidth]{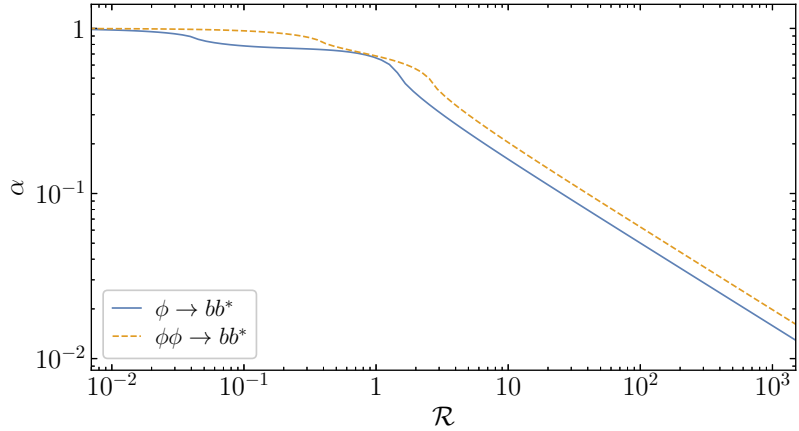}
    \caption{Kinematic factor for the oscillation-averaged decay rates (\ref{eq:phitobsusy2}) for the supersymmetric model (\ref{eq:LlHsusy}). }
    \label{fig:KinSusyK2}
\end{figure}

The computation of the decay rates for the supersymmetric scenario discussed in Section~\ref{sec:susy} follows immediately from the previous discussion. The main difference corresponds to the two terms that source the effective mass of bosons, proportional to $Y^2\phi^2$ and $Y\partial_{\Phi}F(\phi)$, as shown in Eq.~(\ref{eq:meffsusy}). For $k=2$, $\partial_{\Phi}F(\phi)=m_{\phi}\phi/\sqrt{2}$. In this case it is convenient to write
\beq
\mathcal{R} \;=\; \frac{Y^2 \phi_0^2}{2 m_{\phi}^2}\,,
\eeq
so that the corresponding oscillation-averaged phase-space factor, which we denote simply by the corresponding channel, is given by
\begin{align}
\alpha_{\phi\rightarrow bb^*}(\mathcal{R}) \;&=\; \left\langle \left(1-4\mathcal{R}\mathcal{P}^2 \mp 4\sqrt{\mathcal{R}}\mathcal{P} \right)^{1/2}_+ \right\rangle\,,\\
\alpha_{\phi\phi\rightarrow bb^*}(\mathcal{R}) \;&=\; \left\langle \left(1-\mathcal{R}\mathcal{P}^2 \mp \sqrt{\mathcal{R}}\mathcal{P} \right)^{1/2}_+ \right\rangle\,,
\end{align}
(only the first harmonic contributes). Fig.~\ref{fig:KinSusyK2} shows the magnitude of this suppression factor, as the continuous blue curve for the decay process, and the dashed yellow curve for the scattering channel.

\begin{figure}[!ht]
\centering
    \includegraphics[width=0.65\textwidth]{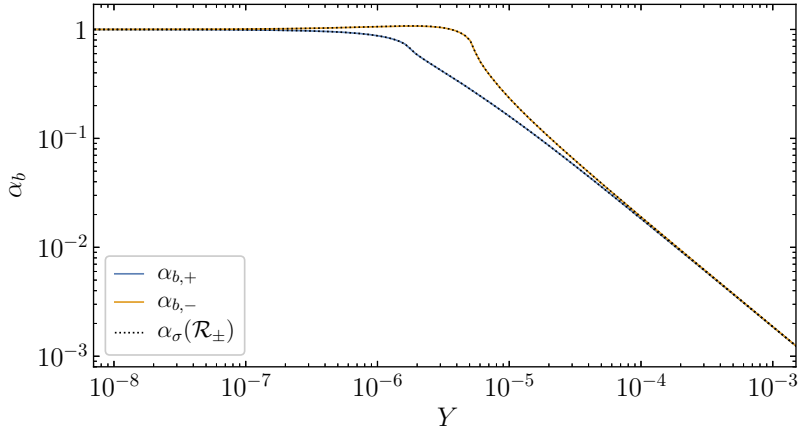}
    \caption{Kinematic factor for the oscillation-averaged decay rates (\ref{eq:phitobsusy4}) for the supersymmetric model (\ref{eq:LlHsusy}) with $k=4$. Here the T-attractor value $\lambda\simeq 3.3\times 10^{-12}$ has been chosen for definiteness.}
    \label{fig:KinSusyK4}
\end{figure}

For $k=4$, $\partial_{\Phi}F(\phi)=\sqrt{\lambda}\phi^2$. Therefore, only the scattering process for bosons is present at lowest order. In this scenario, with the identification $\sigma\rightarrow Y(Y\pm 2\sqrt{\lambda})/2$ in (\ref{eq:kincond}), we write
\beq
\mathcal{R}_{\pm} \;\simeq\; 1.4 \frac{Y}{\lambda}\left(Y\pm 2\sqrt{\lambda}\right)\,.
\eeq
and 
\beq\label{eq:apmm}
\alpha_{b,\pm} \;=\; 8 \sum_{n=1}^{\infty} n| \left(\mathcal{P}^2\right)_n |^2 \left\langle \left( 1-\frac{\mathcal{R_{\pm}}}{n^2}\mathcal{P}^2 \right)^{1/2}_+ \right\rangle\,.
\eeq
Note that, unlike previous cases, here $\mathcal{R}_{\pm}$ can be positive or negative, depending on the magnitude of $Y$ relative to $2\sqrt{\lambda}$. For the T-attractor, $\sqrt{\lambda}\simeq 1.8\times 10^{-6}$. In Fig.~\ref{fig:KinSusyK4} the result of the numerical evaluation of (\ref{eq:apmm}) for the T-attractor is presented, as a function of $Y$. We note that, for $\alpha_{b,+}$, only a suppression in the rate is observed. Shown also in the figure are the values for $\alpha_{\sigma}(\mathcal{R}_{+})$, which are identical to those for $\alpha_{b,+}$. On the other hand, for $\alpha_{b,-}$ an enhancement $<\mathcal{O}(1)$ in the rate is observed for $Y<2\sqrt{\lambda}$, which corresponds to the range in which $\mathcal{R}_-$ is negative. For larger values of the coupling, the decay rate of $\phi$ is suppressed. In Fig.~\ref{fig:KinSusyK4} the oscillation average of $\alpha_{\sigma}(\mathcal{R}_-)$ is also shown, assuming it is equal to 1 for negative $\mathcal{R}_-$. This quantity matches the behavior of $\alpha_{b,-}$.

%%%%%%%%%%%%%%%%%%%%%%%%. e-fold.  %%%%%%%%%%%%%%%%%%

\section{T-attractor inflation}\label{sec:efold}

The rates at which energy densities and temperatures change during reheating depend not only on the shape of the potential (\ref{eq:Vphi}), parametrized by $k$, but also on its normalization, parametrized by $\lambda$, which in turn is determined by the amplitude of the power spectrum of curvature fluctuations. Moreover, the initial condition for $\rho_{\phi}$ depends on the value of the scalar field $\phi$ at the end of inflation. In this Appendix we determine the boundary conditions for $\lambda$ and $\rho_{\rm end}$ under the assumption that the scalar potential responsible for inflation is of the T-model form (\ref{eq:attractor}). 

Inflation is defined as a period of accelerated expansion. Its end is therefore defined as the condition $\ddot{a}=0$, which can be shown to be equivalent to $\dot{\phi}_{\rm end}^2=V(\phi)$~\cite{Ellis:2015pla}. An approximate solution for these conditions for arbitrary $k$ is given by
\beq
\phi_{\rm end} \;=\;\sqrt{\frac{3}{8}}\, M_P \ln\left[ \frac{1}{2} + \frac{k}{3}\left(k+\sqrt{k^2+3}\right) \right]\,.
\eeq
The energy density is in turn determined as $\rho_{\rm end}=\frac{3}{2}V(\phi_{\rm end})$. 
The normalization of the potential can 
be determined from the inflationary slow role parameters as follows. Given the potential in Eq.~(\ref{eq:attractor}),
the slow roll parameters, $\epsilon$ and $\eta$ are
\beq
\epsilon \; \equiv \; \frac{1}{2} M_{P}^2 \left( \frac{V'}{V} \right)^2  \;=\; \frac{k^2}{3} {\rm csch}^2\left(\sqrt{\frac23}\frac{\phi}{M_P}\right) \, ,
\eeq
and 
\beq
 \eta \; \equiv \; M_{P}^2 \left( \frac{V''}{V} \right) \;=\; 
 \frac23 k \left[ k-\cosh\left(\sqrt{\frac23}\frac{\phi}{M_P} \right) \right] {\rm csch}^2\left(\sqrt{\frac23}\frac{\phi}{M_P}\right) \, .
\eeq
The number of $e$-folds between the exit of the horizon of the scale $k_*$ at $\phi_*$, and the end of inflation at $\phi_{\rm end}$, can be computed in the slow-roll approximation as follows,
\beq
N_* \;\simeq\; \frac{1}{M_P^2} \int_{\phi_{\rm end}}^{\phi_*} \frac{V(\phi)}{V'(\phi)}\,d\phi
 \; \simeq \;  \int^{\phi_*}_{\phi_{\rm{end}}} \frac{1}{\sqrt{2 \epsilon}} \frac{d \phi}{M_P}
  \; \simeq \; \frac{3}{2 k} \cosh \left(\sqrt{\frac23}\frac{\phi_*}{M_P} \right) \, .
\eeq
In the slow-roll approximation, the scalar tilt $n_s$ and the tensor-to-scalar ratio $r$ are given by the following expressions,
\begin{align}
n_s \;&\simeq\; 1-6\epsilon_* + 2\eta_* \;=\; 1 - \frac{2k^2(4N_*+3)}{4k^2N_*^2-9} \;\simeq\; 1-\frac{2}{N_*} - \frac{3}{2N_*^2}\,,\\
r \;&\simeq\; 16\epsilon_* \;=\; \frac{48k^2}{4k^2N_*^2-9} \;\simeq\; \frac{12}{N_*^2}\,.
\end{align}
Finally, the amplitude of the curvature power spectrum 
can be expressed as
\beq
A_{S*} \;\simeq\; \frac{V_*}{24 \pi^2 \epsilon_* M_{P}^4 } \;\simeq\; \frac{6^{\frac{k}{2}}}{8k^2\pi^2}\lambda \sinh^2\left(\sqrt{\frac23}\frac{\phi_*}{M_P} \right) \tanh^k\left(\frac{\phi_*}{\sqrt{6} M_P} \right)\, ,
\label{As}
\eeq
where $V_* = V(\phi_*)$.
For the Planck pivot scale $k_*=0.05\,{\rm Mpc}^{-1}$, $\ln(10^{10}A_{S*}) = 3.044$~\cite{Aghanim:2018eyx,planckinf}.
Thus to determine the normalization of the potential given by $\lambda$, we must first obtain $\phi_*$ through $N_*$. 
A good approximation is given by
\beq\label{eq:lambdavsNs}
\lambda \;\simeq\; \frac{18\pi^2 A_{S*}}{6^{k/2}N_*^2}\,.
\eeq
which can be obtained by substitution of $\phi_*(N_*)$ in the expression for $A_{S*}$ and is good to 3\% for $N_*\in(50,60)$.

The number of $e$-folds, $N_*$, can be computed in a self-consistent way assuming there is no entropy production between the end of reheating and the reentry to the horizon of the scale $k_*$ in the radiation or matter-dominated eras. The value of $N_*$ depends on the energy scale of inflation and the duration of reheating, as measured by the deviation of the total equation-of-state parameter $w$ from its value during radiation domination, $w=1/3$. More precisely,~\cite{Martin:2010kz,Liddle:2003as},
\begin{align} \notag
N_* \;=\; \ln&\left[\frac{1}{\sqrt{3}}\left(\frac{\pi^2}{30}\right)^{1/4}\left(\frac{43}{11}\right)^{1/3}\frac{T_0}{H_0}\right] - \ln\left(\frac{k_*}{a_0H_0}\right) + \frac{1}{4}\ln\left(\frac{V_*^2}{M_P^4\rho_{\rm end}}\right)\\ \label{eq:Ns}
&+ \frac{1-3w_{\rm int}}{12(1+w_{\rm int})}\ln\left(\frac{\rho_{\rm rad}}{\rho_{\rm end}}\right) - \frac{1}{12}\ln g_{\rm reh}\,.
\end{align}
The present photon temperature and Hubble parameter, as determined from CMB observations, are given by $T_0=2.7255\,{\rm K}$ and $H_0=67.36\,{\rm km}\,{\rm s}^{-1}\,{\rm Mpc}^{-1}$, respectively~\cite{Aghanim:2018eyx,Fixsen:2009ug}. The scale factor at the present time is normalized as $a_0=1$. The energy at the beginning of the $w=1/3$ era is denoted by $\rho_{\rm rad}$. Note that in general $\rho_{\rm reh}\geq \rho_{\rm rad}$, as $1/3\geq w\geq w_{\phi}$ at inflaton-radiation equality. Finally, $w_{\rm int}$ denotes the $e$-fold average of the equation-of-state parameter during reheating,
\beq
w_{\rm int} \;\equiv\; \frac{1}{N_{\rm rad}-N_{\rm end}} \int_{N_{\rm end}}^{N_{\rm rad}}w(n)\,dn\,.
\eeq

The solution of (\ref{eq:Ns}) for $N_*$ must be found numerically in general. We solve it by iteration. Namely,
using Eq.~(\ref{eq:lambdavsNs}) to determine $\lambda$ and $V_*$ in (\ref{eq:Ns}) allows for a simple solution for $N_*$ if one approximates $w_{\rm int}\approx w_{\phi}$ and $\rho_{\rm rad}\approx \rho_{\rm reh}$, the later given by
$T_{\rm RH}$.\footnote{This is the approximation used in~\cite{Garcia:2020eof,Maity:2018exj}.} One can then substitute into (\ref{As}) and use it as a boundary condition for the numerical solution of the system (\ref{eq:rhoeom1})-(\ref{hub}), which permits a better determination of $w_{\rm int}$ and $\rho_{\rm rad}$, and hence of $N_*$.\footnote{Since $w\rightarrow 1/3$ asymptotically, a threshold for the beginning of the radiation dominated era must be chosen. We chose it here as $|w-1/3|=10^{-2}$.} This method converges rapidly, to the second decimal place after only one iteration. 

A few features are common to all inflaton depletion processes. 
For $k=4$, $w_{\phi}=1/3$, and the number of $e$-folds is independent on the details of reheating and is always equal to 55.9 for T-attractor inflation. Note also that for any value of $k$ the CMB parameters $n_s$ and $r$ converge to the same attractor limit at $N_*\gg 1$. This feature allows the identification of a domain of $e$-folds compatible with Planck data combined with BICEP2/Keck results~\cite{{Aghanim:2018eyx,planckinf}}. At 68\%, $45.2\lesssim N_*\lesssim 76.7$, while at 95\%, $49.7\lesssim N_*\lesssim 66.7$.

\begin{figure}[!ht]
\centering
    \includegraphics[width=0.65\textwidth]{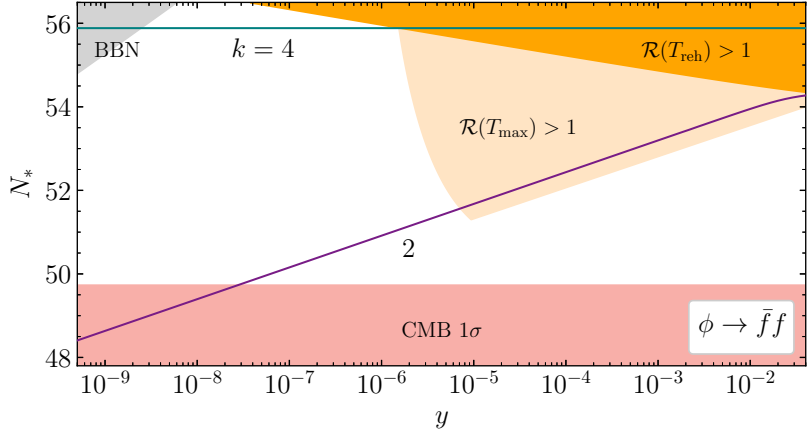}
    \caption{Number of $e$-folds from the exit of the Planck pivot scale to the end of inflation, as a function of $k$ and $y$, for the decay of $\phi$ into fermions. Here the Standard Model value $g_{\rm reh}=427/4$ is used. The gray region is incompatible with Big Bang Nucleosynthesis. The light red region is disfavored at 68\% by Planck+BKP.}
    \label{fig:NsF}
\end{figure}

Fig.~\ref{fig:NsF} shows the numerical solution for $N_*$ for the perturbative decay of the inflaton into fermions for $k=2,4$. The whole parameter space depicted there lies within the Planck+BICEP2/Keck 95\% CL region for the scalar tilt $n_s$ at low tensor-to-scalar ratio. The 68\% CL exclusion region is shown in light red. The excluded region in gray corresponds to reheating temperatures lower than $1\,{\rm MeV}$, incompatible with big bang nucleosynthesis~\cite{Hasegawa:2019jsa,Fields:2019pfx}. The region in light orange corresponds to values of the Yukawa coupling for which a kinematic suppression is present until some time $t_{\mathcal{R}}\geq t_{\rm max}$ (c.f.~Eq.~(\ref{eq:aRdef})). In the orange region, this kinematic suppression is present until the end of reheating, and therefore the approximation $m_{\eff}=0$ leads to an inadequate estimate for the temperature of the inflaton decay products throughout the duration of reheating. We note that, for $k=4$, $w_{\phi}=1/3$, and therefore the number of $e$-folds is $N_*\simeq 55.9$, independently of the decay rate. We do not show the values of $N_*$ for $k=6$, as in that case the kinematic suppression leads to $T_{\rm RH}<1\,{\rm MeV}$ for $y\lesssim 5\times 10^{-2}$. At larger values of the coupling we expect our approximations to break down.

\begin{figure}[!ht]
\centering
    \includegraphics[width=0.65\textwidth]{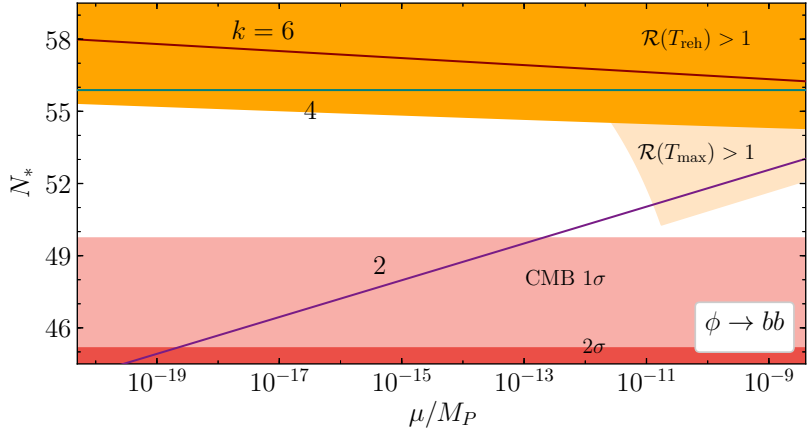}
    \caption{Number of $e$-folds from the exit of the Planck pivot scale to the end of inflation, as a function of $k$ and $\mu$, for the decay of $\phi$ into bosons. Here the Standard Model value $g_{\rm reh}=427/4$ is used. The light red (red) region is disfavored at 68\% (95\%) by Planck+BKP.}
    \label{fig:NsB}
\end{figure}

Fig.~\ref{fig:NsB} shows the numerically calculated number of $e$-folds for the Planck pivot scale for the process $\phi\rightarrow bb$, for $k=2,4,6$. In this case, the light red shaded region is excluded by CMB observations to 68\% CL, while the red region is excluded at 95\%. In the light orange area, a kinematic {\em enhancement} of the decay rate is present from $t_{\rm end}$ to $t_{\mathcal{R}}\geq t_{\rm max}$. In the orange region, the enhancement is present until the end of reheating. For $k=6$ this reduces the value of $N_*$, albeit only at the $\mathcal{O}(1\%)$ level. For $k=4$, $N_*\simeq 55.9$.

\begin{figure}[!t]
\centering
    \includegraphics[width=0.65\textwidth]{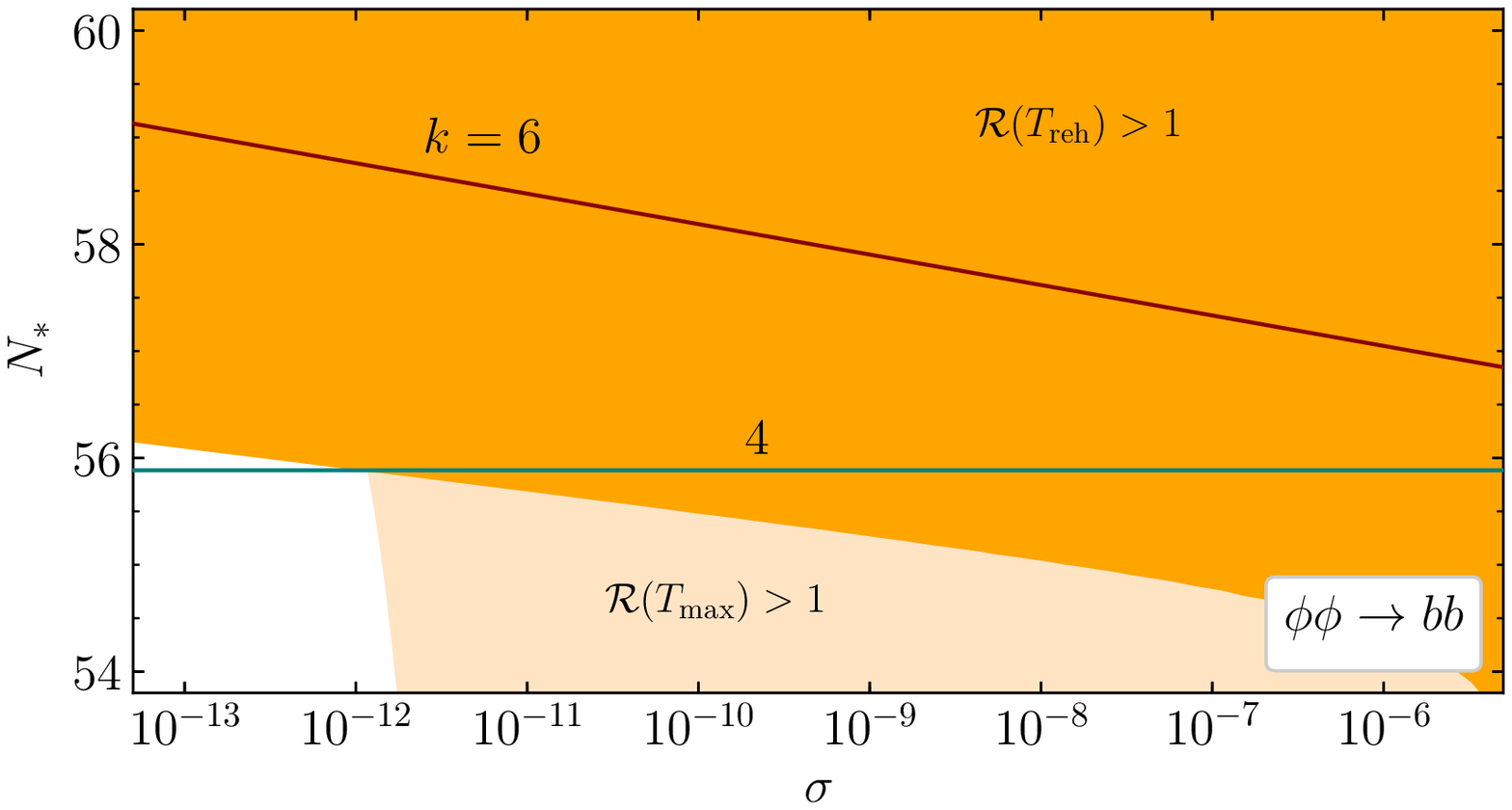}
    \caption{Number of $e$-folds from the exit of the Planck pivot scale to the end of inflation, as a function of $k$ and $\sigma$, for the scattering of $\phi$ into bosons. Here the Standard Model value $g_{\rm reh}=427/4$ is used.}
    \label{fig:NsB2}
\end{figure}

Finally the numerical results for the scattering process $\phi\phi\rightarrow bb$ are shown in Fig.~\ref{fig:NsB2}.
Recall however, that by itself, this process cannot reheat the universe unless $k>3$ (see Table~\ref{Tab:table}). For both $k = 4$ and $k=6$, for couplings,
$\sigma > 10^{-12}$, we are always in a regime where ${\cal R}(T_{\rm RH}) > 1$.

These results can be used to fix the potential normalization, $\lambda$ and $\rho_{\rm end}$.
As one can see from Figs.~\ref{fig:NsF} and \ref{fig:NsB}, the numerically determined value of $N_*$ depends on the couplings $y$ and $\mu$ for $k=2$.  For $\mu \la m_\phi y$, decays are dominated by fermionic final states and
for $k=2$, 
\beq 
\lambda \;\simeq\; 2.5 \times 10^{-11}\left(\frac{10^{-7}}{y}\right)^{1/80}\,,  \qquad \rho_{\rm end} \;\simeq\; \left(5.2 \times 10^{15}  {\rm GeV}\right)^4\left(\frac{10^{-7}}{y}\right)^{1/80} \, .
\label{lambdarho2}
\eeq
 When $\mu > m_\phi y$, one can place $y$ with $\mu/m_\phi$ in (\ref{lambdarho2}) to obtain $\lambda$ and $\rho_{\rm end}$. 
For $k=4$, $N_*$ does not depend on the choice of couplings, and we have simply
\beq 
\lambda \;\simeq\; 3.3 \times 10^{-12}\,, \qquad \rho_{\rm end} \;\simeq\; \left(4.8 \times 10^{15} {\rm GeV}\right)^4 \, .
\eeq

\bibliographystyle{apsrev4-1}

%\bibliography{biblio} 

\begin{thebibliography}{99}  




%%%%%%%%%%%%%%%%%%%    INTRODUCTION    %%%%%%%%%%%%%%%%%%%%%%%%%%%%%%%

 \bibitem{reviews}
   K.~A.~Olive,
  %``Inflation,''
  Phys.\ Rept.\  {\bf 190} (1990) 307;
  %%CITATION = PRPLC,190,307;%%
A. D. Linde, {\it Particle  
Physics and
Inflationary Cosmology} (Harwood, Chur, Switzerland, 1990); 
  D.~H.~Lyth and A.~Riotto,
%``Particle physics models of inflation and the cosmological density  perturbation,''
{\it Phys.\ Rep.}  {\bf 314} (1999) 1
[arXiv:hep-ph/9807278];
%\cite{Linde:2000kn}
%\bibitem{Linde:2000kn}
A.~D.~Linde,
%``Inflationary cosmology,''
Phys. Rept. \textbf{333}, 575-591 (2000);
%doi:10.1016/S0370-1573(00)00038-7
%33 citations counted in INSPIRE as of 20 Jul 2020
J.~Martin, C.~Ringeval and V.~Vennin,
  %``EncyclopÔøΩdia Inflationaris,''
  Phys.\ Dark Univ.\  {\bf 5-6}, 75-235 (2014)
  [arXiv:1303.3787 [astro-ph.CO]];
  %%CITATION = ARXIV:1303.3787;%%
  J.~Martin, C.~Ringeval, R.~Trotta and V.~Vennin,
  %``The Best Inflationary Models After Planck,''
  JCAP {\bf 1403} (2014) 039
  [arXiv:1312.3529 [astro-ph.CO]];
  %%CITATION = ARXIV:1312.3529;%%
  %46 citations counted in INSPIRE as of 23 May 2015
 J.~Martin,
  %``The Observational Status of Cosmic Inflation after Planck,''
  Astrophys.\ Space Sci.\ Proc.\  {\bf 45}, 41 (2016)
 % doi:10.1007/978-3-319-44769-8_2
  [arXiv:1502.05733 [astro-ph.CO]].
  %%CITATION = doi:10.1007/978-3-319-44769-8_2;%%

%\cite{Akrami:2018odb}
\bibitem{planckinf}
Y.~Akrami \textit{et al.} [Planck],
%``Planck 2018 results. X. Constraints on inflation,''
Astron. Astrophys. \textbf{641}, A10 (2020)
%doi:10.1051/0004-6361/201833887
[arXiv:1807.06211 [astro-ph.CO]].
%1080 citations counted in INSPIRE as of 03 Dec 2020


\bibitem{dl}  
A.~Dolgov and A.~D.~Linde,
%``Baryon Asymmetry in Inflationary Universe,''
Phys. Lett. B \textbf{116}, 329 (1982);
%doi:10.1016/0370-2693(82)90292-1
%499 citations counted in INSPIRE as of 28 Jun 2020
L.~Abbott, E.~Farhi and M.~B.~Wise,
%``Particle Production in the New Inflationary Cosmology,''
Phys. Lett. B \textbf{117}, 29 (1982).
%doi:10.1016/0370-2693(82)90867-X
%581 citations counted in INSPIRE as of 28 Jun 2020


%\cite{Nanopoulos:1983up}
\bibitem{nos}
D.~V.~Nanopoulos, K.~A.~Olive and M.~Srednicki,
%``After Primordial Inflation,''
Phys. Lett. B \textbf{127}, 30-34 (1983);
%doi:10.1016/0370-2693(83)91624-6
%282 citations counted in INSPIRE as of 28 Jun 2020


\bibitem{Giudice:2000ex}
  G.~F.~Giudice, E.~W.~Kolb and A.~Riotto,
  %``Largest temperature of the radiation era and its cosmological implications,''
  Phys.\ Rev.\ D {\bf 64} (2001) 023508
 % doi:10.1103/PhysRevD.64.023508
  [hep-ph/0005123];
  %%CITATION = doi:10.1103/PhysRevD.64.023508;%%
  %319 citations counted in INSPIRE as of 04 Apr 2018
   D.~J.~H.~Chung, E.~W.~Kolb and A.~Riotto,
  %``Production of massive particles during reheating,''
  Phys.\ Rev.\ D {\bf 60} (1999) 063504
%  doi:10.1103/PhysRevD.60.063504
  [hep-ph/9809453].
  %%CITATION = doi:10.1103/PhysRevD.60.063504;%%
  %270 citations counted in INSPIRE as of 04 Apr 2018
  
    \bibitem{grav2}
  E.~Dudas, Y.~Mambrini and K.~Olive,
  %``Case for an EeV Gravitino,''
  Phys.\ Rev.\ Lett.\  {\bf 119} (2017) no.5,  051801
%  doi:10.1103/PhysRevLett.119.051801
  [arXiv:1704.03008 [hep-ph]].
  %%CITATION = doi:10.1103/PhysRevLett.119.051801;%%
  %10 citations counted in INSPIRE as of 17 Feb 2018
  
  \bibitem{Garcia:2017tuj}
M.~A.~G.~Garcia, Y.~Mambrini, K.~A.~Olive and M.~Peloso,
%``Enhancement of the Dark Matter Abundance Before Reheating: Applications to Gravitino Dark Matter,''
Phys. Rev. D \textbf{96}, no.10, 103510 (2017)
%doi:10.1103/PhysRevD.96.103510
[arXiv:1709.01549 [hep-ph]].
%32 citations counted in INSPIRE as of 07 Jun 2020

%\cite{Chen:2017kvz}
\bibitem{Chen:2017kvz}
S.~L.~Chen and Z.~Kang,
%``On UltraViolet Freeze-in Dark Matter during Reheating,''
JCAP \textbf{05}, 036 (2018)
%doi:10.1088/1475-7516/2018/05/036
[arXiv:1711.02556 [hep-ph]].
%20 citations counted in INSPIRE as of 03 Dec 2020

 \bibitem{Garcia:2020eof}
M.~A.~Garcia, K.~Kaneta, Y.~Mambrini and K.~A.~Olive,
%``Reheating and Post-inflationary Production of Dark Matter,''
Phys. Rev. D \textbf{101} (2020) no.12, 123507
%doi:10.1103/PhysRevD.101.123507
[arXiv:2004.08404 [hep-ph]].
%3 citations counted in INSPIRE as of 20 May 2020

%\cite{Bernal:2020gzm}
\bibitem{Bernal:2020gzm}
N.~Bernal,
%``Boosting Freeze-in through Thermalization,''
JCAP \textbf{10}, 006 (2020)
%doi:10.1088/1475-7516/2020/10/006
[arXiv:2005.08988 [hep-ph]].
%12 citations counted in INSPIRE as of 04 Dec 2020

%\cite{Co:2020xaf}
\bibitem{Co:2020xaf}
R.~T.~Co, E.~Gonzalez and K.~Harigaya,
%``Increasing Temperature toward the Completion of Reheating,''
JCAP \textbf{11}, 038 (2020)
%doi:10.1088/1475-7516/2020/11/038
[arXiv:2007.04328 [astro-ph.CO]].
%2 citations counted in INSPIRE as of 04 Dec 2020


\bibitem{Davidson:2000er} 
  S.~Davidson and S.~Sarkar,
  %``Thermalization after inflation,''
  JHEP {\bf 0011}, 012 (2000)
  [hep-ph/0009078].
  %%CITATION = HEP-PH/0009078;%%

  \bibitem{Harigaya:2013vwa}
  K.~Harigaya, K.~Mukaida and M.~Yamada,
%``Dark Matter Production during the Thermalization Era,''
JHEP \textbf{07} (2019), 059
%doi:10.1007/JHEP07(2019)059
[arXiv:1901.11027 [hep-ph]];
%9 citations counted in INSPIRE as of 09 Jun 2020
K.~Harigaya, M.~Kawasaki, K.~Mukaida and M.~Yamada,
%``Dark Matter Production in Late Time Reheating,''
Phys. Rev. D \textbf{89} (2014) no.8, 083532
%doi:10.1103/PhysRevD.89.083532
[arXiv:1402.2846 [hep-ph]];
%54 citations counted in INSPIRE as of 09 Jun 2020 
K.~Harigaya and K.~Mukaida,
%``Thermalization after/during Reheating,''
JHEP \textbf{05}, 006 (2014)
%doi:10.1007/JHEP05(2014)006
[arXiv:1312.3097 [hep-ph]].
%77 citations counted in INSPIRE as of 03 Dec 2020

%\cite{Mukaida:2015ria}
\bibitem{Mukaida:2015ria}
K.~Mukaida and M.~Yamada,
%``Thermalization Process after Inflation and Effective Potential of Scalar Field,''
JCAP \textbf{02}, 003 (2016)
%doi:10.1088/1475-7516/2016/02/003
[arXiv:1506.07661 [hep-ph]].
%48 citations counted in INSPIRE as of 03 Dec 2020

 
  
\bibitem{GA}
 M.~A.~G.~Garcia and M.~A.~Amin,
  %``Prethermalization production of dark matter,''
  Phys.\ Rev.\ D {\bf 98}, no. 10, 103504 (2018)
 % doi:10.1103/PhysRevD.98.103504
  [arXiv:1806.01865 [hep-ph]];
  %%CITATION = doi:10.1103/PhysRevD.98.103504;%%
  
  \bibitem{Bernal:2019mhf}
N.~Bernal, F.~Elahi, C.~Maldonado and J.~Unwin,
%``Ultraviolet Freeze-in and Non-Standard Cosmologies,''
JCAP \textbf{11} (2019), 026
%doi:10.1088/1475-7516/2019/11/026
[arXiv:1909.07992 [hep-ph]].
%22 citations counted in INSPIRE as of 01 Nov 2020


\bibitem{Kallosh:2013hoa}
R.~Kallosh and A.~Linde,
%``Universality Class in Conformal Inflation,''
JCAP \textbf{07} (2013), 002
%doi:10.1088/1475-7516/2013/07/002
[arXiv:1306.5220 [hep-th]].

\bibitem{gravitino}
  H.~Pagels and J.~R.~Primack,
  %``Supersymmetry, Cosmology and New TeV Physics,''
  Phys.\ Rev.\ Lett.\  {\bf 48}, 223 (1982).
 % doi:10.1103/PhysRevLett.48.223
  %%CITATION = doi:10.1103/PhysRevLett.48.223;%%
  
  \bibitem{ehnos}
  J.~Ellis, J.~Hagelin, D.~Nanopoulos, K.~Olive and M.~Srednicki,
                Nucl.\ Phys.\ B {\bf 238} (1984) 453.
                %%CITATION = NUPHA,B238,453;%%
                
                \bibitem{kl}
  M.~Y.~Khlopov and A.~D.~Linde,
  %``Is It Easy to Save the Gravitino?,''
  Phys.\ Lett.\ B {\bf 138}, 265 (1984).
  %%CITATION = PHLTA,B138,265;%%     
  
  \bibitem{oss}
  K.~A.~Olive, D.~N.~Schramm and M.~Srednicki,
  %``Gravitinos as the Cold Dark Matter in an omega = 1 Universe,''
  Nucl.\ Phys.\ B {\bf 255}, 495 (1985).
 % doi:10.1016/0550-3213(85)90149-X
  %%CITATION = doi:10.1016/0550-3213(85)90149-X;%%
  
  %%%%%%%%%%%%%%%%%%%%%.  FIMP   %%%%%%%%%%%%%%%%%%%%%%

\bibitem{fimp}
  L.~J.~Hall, K.~Jedamzik, J.~March-Russell and S.~M.~West,
  %``Freeze-In Production of FIMP Dark Matter,''
  JHEP {\bf 1003} (2010) 080
%  doi:10.1007/JHEP03(2010)080
  [arXiv:0911.1120 [hep-ph]];
  %%CITATION = doi:10.1007/JHEP03(2010)080;%%
  %182 citations counted in INSPIRE as of 25 Oct 2016
  X.~Chu, T.~Hambye and M.~H.~G.~Tytgat,
  %``The Four Basic Ways of Creating Dark Matter Through a Portal,''
  JCAP {\bf 1205} (2012) 034
%  doi:10.1088/1475-7516/2012/05/034
  [arXiv:1112.0493 [hep-ph]];
  X.~Chu, Y.~Mambrini, J.~Quevillon and B.~Zaldivar,
  %``Thermal and non-thermal production of dark matter via Z'-portal(s),''
  JCAP {\bf 1401} (2014) 034
%  doi:10.1088/1475-7516/2014/01/034
  [arXiv:1306.4677 [hep-ph]];
  A.~Biswas, D.~Borah and A.~Dasgupta,
%``UV complete framework of freeze-in massive particle dark matter,''
Phys.\ Rev.\ D \textbf{99}, no.1, 015033 (2019)
% doi:10.1103/PhysRevD.99.015033
[arXiv:1805.06903 [hep-ph]].

\bibitem{Bernal}
N.~Bernal, J.~Rubio and H.~Veermäe,
%``UV Freeze-in in Starobinsky Inflation,''
[arXiv:2006.02442 [hep-ph]].
%0 citations counted in INSPIRE as of 06 Jun 2020
%\bibliography{biblio} 

  
  \bibitem{Bernal:2017kxu}
  N.~Bernal, M.~Heikinheimo, T.~Tenkanen, K.~Tuominen and V.~Vaskonen,
  %``The Dawn of FIMP Dark Matter: A Review of Models and Constraints,''
  Int.\ J.\ Mod.\ Phys.\ A {\bf 32} (2017) no.27,  1730023
 % doi:10.1142/S0217751X1730023X
  [arXiv:1706.07442 [hep-ph]].
  %%CITATION = doi:10.1142/S0217751X1730023X;%%
  %29 citations counted in INSPIRE as of 17 Feb 2018
  


  
   \bibitem{ekn}
  J.~R.~Ellis, J.~E.~Kim and D.~V.~Nanopoulos,
  %``Cosmological Gravitino Regeneration and Decay,''
  Phys.\ Lett.\ B {\bf 145}, 181 (1984).
  %%CITATION = PHLTA,B145,181;%%         

    
  
  \bibitem{Juszkiewicz:gg}
R.~Juszkiewicz, J.~Silk and A.~Stebbins,
%``Constraints On Cosmologically Regenerated Gravitinos,''
Phys.\ Lett.\ B {\bf 158} (1985) 463.
%%CITATION = PHLTA,B158,463;%%

\bibitem{mmy}
T.~Moroi, H.~Murayama and M.~Yamaguchi,
  %``Cosmological constraints on the light stable gravitino,''
  Phys.\ Lett.\ B {\bf 303}, 289 (1993).
  %%CITATION = PHLTA,B303,289;%%

 
  
\bibitem{Kawasaki:1994af} 
  M.~Kawasaki and T.~Moroi,
  %``Gravitino production in the inflationary universe and the effects on big bang nucleosynthesis,''
  Prog.\ Theor.\ Phys.\  {\bf 93}, 879 (1995)
  [hep-ph/9403364].
  %%CITATION = HEP-PH/9403364,;%%
  
   \bibitem{Moroi:1995fs} 
  T.~Moroi,
  %``Effects of the gravitino on the inflationary universe,''
  hep-ph/9503210.
  %%CITATION = HEP-PH/9503210;%%
  
  \bibitem{enor}
  J.~R.~Ellis, D.~V.~Nanopoulos, K.~A.~Olive and S.~J.~Rey,
  %``On the thermal regeneration rate for light gravitinos in the early universe,''
  Astropart.\ Phys.\  {\bf 4}, 371 (1996)
  [hep-ph/9505438].
  %%CITATION = HEP-PH/9505438;%%
  
      \bibitem{Giudice:1999am} 
  G.~F.~Giudice, A.~Riotto and I.~Tkachev,
  %``Thermal and nonthermal production of gravitinos in the early universe,''
  JHEP {\bf 9911}, 036 (1999)
  [hep-ph/9911302].
  %%CITATION = HEP-PH/9911302;%%
  
   \bibitem{bbb}
   M.~Bolz, A.~Brandenburg and W.~Buchmuller,
  %``Thermal production of gravitinos,''
  Nucl.\ Phys.\ B {\bf 606}, 518 (2001)
  [Erratum-ibid.\ B {\bf 790}, 336 (2008)]
  [hep-ph/0012052].
  %%CITATION = HEP-PH/0012052;%%
  
  \bibitem{cefo}
 R.~H.~Cyburt, J.~Ellis, B.~D.~Fields and K.~A.~Olive,
 %"Updated Nucleosynthesis Constraints on Unstable Relic Particles"%
 Phys.\ Rev.\ D {\bf 67}, 103521 (2003) [astro-ph/0211258].
 %%%%


   \bibitem{kmy}
 K.~Kohri, T.~Moroi and A.~Yotsuyanagi,
   %``Big-bang nucleosynthesis with unstable gravitino and upper bound on the
  %reheating temperature,''
  Phys.\ Rev.\ D {\bf 73}, 123511 (2006)
  [arXiv:hep-ph/0507245].
  %%CITATION = HEP-PH 0507245;%%

  
  
  \bibitem{stef}
    F.~D.~Steffen,
  %``Gravitino dark matter and cosmological constraints,''
  JCAP {\bf 0609}, 001 (2006)
  [arXiv:hep-ph/0605306].
  %%CITATION = JCAPA,0609,001;%%
  
  \bibitem{Pradler:2006qh} 
  J.~Pradler and F.~D.~Steffen,
  %``Thermal gravitino production and collider tests of leptogenesis,''
  Phys.\ Rev.\ D {\bf 75}, 023509 (2007)
  [hep-ph/0608344].
  %%CITATION = HEP-PH/0608344;%%
  
  \bibitem{ps2}
   J.~Pradler and F.~D.~Steffen,
  %``Constraints on the Reheating Temperature in Gravitino Dark Matter Scenarios,''
  Phys.\ Lett.\ B {\bf 648}, 224 (2007)
  [hep-ph/0612291].
  %%CITATION = HEP-PH/0612291;%%
  
   \bibitem{kkmy}
  M.~Kawasaki, K.~Kohri, T~Moroi and A.Yotsuyanagi,
%"Big-Bang Nucleosynthesis and Gravitino"%
Phys.\ Rev.\ D {\bf 78}, 065011 (2008) [arXiv:0804.3745 [hep-ph]].
  
   \bibitem{rs}
   V.~S.~Rychkov and A.~Strumia,
  %``Thermal production of gravitinos,''
  Phys.\ Rev.\ D {\bf 75}, 075011 (2007)
  [hep-ph/0701104].
  %%CITATION = HEP-PH/0701104;%%

  
    \bibitem{egnop}
  J.~Ellis, M.~A.~G.~Garcia, D.~V.~Nanopoulos, K.~A.~Olive and M.~Peloso,
  %``Post-Inflationary Gravitino Production Revisited,''
  JCAP {\bf 1603}, no. 03, 008 (2016)
%  doi:10.1088/1475-7516/2016/03/008
  [arXiv:1512.05701 [astro-ph.CO]].
  %%CITATION = doi:10.1088/1475-7516/2016/03/008;%%

  
  \bibitem{Eberl:2020fml}
H.~Eberl, I.~D.~Gialamas and V.~C.~Spanos,
%``Gravitino thermal production revisited,''
[arXiv:2010.14621 [hep-ph]].
%2 citations counted in INSPIRE as of 22 Dec 2020
  

 

 



  \bibitem{Benakli:2017whb} 
  K.~Benakli, Y.~Chen, E.~Dudas and Y.~Mambrini,
  %``Minimal model of gravitino dark matter,''
  Phys.\ Rev.\ D {\bf 95}, no. 9, 095002 (2017)
%  doi:10.1103/PhysRevD.95.095002
  [arXiv:1701.06574 [hep-ph]].
  %%CITATION = doi:10.1103/PhysRevD.95.095002;%%
  
 
  
  \bibitem{grav3}
  E.~Dudas, T.~Gherghetta, Y.~Mambrini and K.~A.~Olive,
  %``Inflation and High-Scale Supersymmetry with an EeV Gravitino,''
  Phys.\ Rev.\ D {\bf 96} (2017) no.11,  115032
%  doi:10.1103/PhysRevD.96.115032
  [arXiv:1710.07341 [hep-ph]];
  %%CITATION = doi:10.1103/PhysRevD.96.115032;%%
  %3 citations counted in INSPIRE as of 17 Feb 2018
   %%CITATION = doi:10.1103/PhysRevD.95.095002;%%
  %16 citations counted in INSPIRE as of 17 Feb 2018
 E.~Dudas, T.~Gherghetta, K.~Kaneta, Y.~Mambrini and K.~A.~Olive,
  %``Gravitino decay in high scale supersymmetry with R -parity violation,''
  Phys.\ Rev.\ D {\bf 98}, no. 1, 015030 (2018)
%  doi:10.1103/PhysRevD.98.015030
  [arXiv:1805.07342 [hep-ph]].
  %%CITATION = doi:10.1103/PhysRevD.98.015030;%%
  S.~A.~R.~Ellis, T.~Gherghetta, K.~Kaneta and K.~A.~Olive,
  %``New Weak-Scale Physics from SO(10) with High-Scale Supersymmetry,''
  Phys.\ Rev.\ D {\bf 98}, no. 5, 055009 (2018)
 % doi:10.1103/PhysRevD.98.055009
  [arXiv:1807.06488 [hep-ph]];
  %%CITATION = doi:10.1103/PhysRevD.98.055009;%%
  %\cite{Kaneta:2019yjn}
%\bibitem{Kaneta:2019yjn}
K.~Kaneta, Y.~Mambrini, K.~A.~Olive and S.~Verner,
%``Inflation and Leptogenesis in High-Scale Supersymmetry,''
Phys. Rev. D \textbf{101}, no.1, 015002 (2020)
%doi:10.1103/PhysRevD.101.015002
[arXiv:1911.02463 [hep-ph]].
%4 citations counted in INSPIRE as of 05 Dec 2020



%%%%%%%%%%%%%%%%%%%%%%%. UV Freeze in.    %%%%%%%%%%%%%%%%%%%


 \bibitem{Bhattacharyya:2018evo}
  G.~Bhattacharyya, M.~Dutra, Y.~Mambrini and M.~Pierre,
  %``Freezing-in dark matter through a heavy invisible $Z'$,''
  Phys.\ Rev.\ D {\bf 98} (2018) no.3,  035038
%  doi:10.1103/PhysRevD.98.035038
  [arXiv:1806.00016 [hep-ph]];
  %%CITATION = doi:10.1103/PhysRevD.98.035038;%%
  %2 citations counted in INSPIRE as of 03 Sep 2018 
  A.~Banerjee, G.~Bhattacharyya, D.~Chowdhury and Y.~Mambrini,
%``Dark matter seeping through dynamic gauge kinetic mixing,''
JCAP \textbf{12} (2019), 009
%doi:10.1088/1475-7516/2019/12/009
[arXiv:1905.11407 [hep-ph]].
%5 citations counted in INSPIRE as of 13 May 2020







\bibitem{SO10}
  Y.~Mambrini, K.~A.~Olive, J.~Quevillon and B.~Zaldivar,
  %``Gauge Coupling Unification and Nonequilibrium Thermal Dark Matter,''
  Phys.\ Rev.\ Lett.\  {\bf 110} (2013) no.24,  241306
 % doi:10.1103/PhysRevLett.110.241306
  [arXiv:1302.4438 [hep-ph]];
  N.~Nagata, K.~A.~Olive and J.~Zheng,
  %``Weakly-Interacting Massive Particles in Non-supersymmetric SO(10) Grand Unified Models,''
  JHEP {\bf 1510}, 193 (2015)
%  doi:10.1007/JHEP10(2015)193
  [arXiv:1509.00809 [hep-ph]];
  %%CITATION = doi:10.1007/JHEP10(2015)193;%%
   Y.~Mambrini, N.~Nagata, K.~A.~Olive and J.~Zheng,
  %``Vacuum Stability and Radiative Electroweak Symmetry Breaking in an SO(10) Dark Matter Model,''
  Phys.\ Rev.\ D {\bf 93} (2016) no.11,  111703
 % doi:10.1103/PhysRevD.93.111703
  [arXiv:1602.05583 [hep-ph]];
  X.~Chu, Y.~Mambrini, J.~Quevillon and B.~Zaldivar,
  %``Thermal and non-thermal production of dark matter via Z'-portal(s),''
  JCAP {\bf 1401} (2014) 034
%  doi:10.1088/1475-7516/2014/01/034
  [arXiv:1306.4677 [hep-ph]];
      Y.~Mambrini, N.~Nagata, K.~A.~Olive, J.~Quevillon and J.~Zheng,
  %``Dark matter and gauge coupling unification in nonsupersymmetric SO(10) grand unified models,''
  Phys.\ Rev.\ D {\bf 91} (2015) no.9,  095010
%  doi:10.1103/PhysRevD.91.095010
  [arXiv:1502.06929 [hep-ph]];
   N.~Nagata, K.~A.~Olive and J.~Zheng,
  %``Asymmetric Dark Matter Models in SO(10),''
  JCAP {\bf 1702}, no. 02, 016 (2017)
%  doi:10.1088/1475-7516/2017/02/016
  [arXiv:1611.04693 [hep-ph]].
  %%CITATION = doi:10.1088/1475-7516/2017/02/016;%%



%\cite{Chowdhury:2018tzw}
\bibitem{Chowdhury:2018tzw}
D.~Chowdhury, E.~Dudas, M.~Dutra and Y.~Mambrini,
%``Moduli Portal Dark Matter,''
Phys. Rev. D \textbf{99} (2019) no.9, 095028
%doi:10.1103/PhysRevD.99.095028
[arXiv:1811.01947 [hep-ph]].
%21 citations counted in INSPIRE as of 01 Nov 2020


%\cite{Anastasopoulos:2020gbu}
\bibitem{Anastasopoulos:2020gbu}
P.~Anastasopoulos, K.~Kaneta, Y.~Mambrini and M.~Pierre,
%``Energy-momentum portal to dark matter and emergent gravity,''
Phys. Rev. D \textbf{102} (2020) no.5, 055019
%doi:10.1103/PhysRevD.102.055019
[arXiv:2007.06534 [hep-ph]];
%3 citations counted in INSPIRE as of 01 Nov 2020
P.~Brax, K.~Kaneta, Y.~Mambrini and M.~Pierre,
%``Disformal Dark Matter,''
[arXiv:2011.11647 [hep-ph]];
P.~Anastasopoulos, M.~Bianchi, D.~Consoli and E.~Kiritsis,
%``String (gravi)photons, ''dark brane photons'', holography and the hypercharge portal,''
[arXiv:2010.07320 [hep-ph]].


\bibitem{Bernal:2018qlk}
  N.~Bernal, M.~Dutra, Y.~Mambrini, K.~Olive, M.~Peloso and M.~Pierre,
  %``Spin-2 Portal Dark Matter,''
  Phys.\ Rev.\ D {\bf 97} (2018) no.11,  115020
%  doi:10.1103/PhysRevD.97.115020
  [arXiv:1803.01866 [hep-ph]].
  %%CITATION = doi:10.1103/PhysRevD.97.115020;%%
  %3 citations counted in INSPIRE as of 03 Sep 2018 
  
  
  

%\cite{Bernal:2020fvw}
\bibitem{Bernal:2020fvw}
N.~Bernal, A.~Donini, M.~G.~Folgado and N.~Rius,
%``Kaluza-Klein FIMP Dark Matter in Warped Extra-Dimensions,''
JHEP \textbf{09} (2020), 142
%doi:10.1007/JHEP09(2020)142
[arXiv:2004.14403 [hep-ph]].
%5 citations counted in INSPIRE as of 01 Nov 2020

%\cite{Garcia:2020hyo}
\bibitem{gmov}
M.~A.~G.~Garcia, Y.~Mambrini, K.~A.~Olive and S.~Verner,
%``Case for decaying spin- 3/2 dark matter,''
Phys. Rev. D \textbf{102}, no.8, 083533 (2020)
%doi:10.1103/PhysRevD.102.083533
[arXiv:2006.03325 [hep-ph]].
%7 citations counted in INSPIRE as of 05 Dec 2020


\bibitem{Kaneta:2019zgw}
  K.~Kaneta, Y.~Mambrini and K.~A.~Olive,
  %``Radiative production of nonthermal dark matter,''
  Phys.\ Rev.\ D {\bf 99} (2019) no.6,  063508
%  doi:10.1103/PhysRevD.99.063508
  [arXiv:1901.04449 [hep-ph]].
  %%CITATION = doi:10.1103/PhysRevD.99.063508;%%
  %4 citations counted in INSPIRE as of 21 May 2019

  
 

%%%%%%%%%%%%%%%%%%%%%%% INFLATION.   %%%%%%%%%%%%%%%%%





\bibitem{Starobinsky:1980te}
A.~A.~Starobinsky,
%``A New Type of Isotropic Cosmological Models Without Singularity,''
Adv. Ser. Astrophys. Cosmol. \textbf{3} (1987), 130-133
%doi:10.1016/0370-2693(80)90670-X

\bibitem{Mukhanov:1981xt}
V.~F.~Mukhanov and G.~V.~Chibisov,
%``Quantum Fluctuations and a Nonsingular Universe,''
JETP Lett. \textbf{33} (1981), 532-535

\bibitem{Starobinsky:1983zz}
A.~A.~Starobinsky,
%``The Perturbation Spectrum Evolving from a Nonsingular Initially De-Sitter Cosmology and the Microwave Background Anisotropy,''
Sov. Astron. Lett. \textbf{9} (1983), 302


\bibitem{Kainulainen:2016vzv}
K.~Kainulainen, S.~Nurmi, T.~Tenkanen, K.~Tuominen and V.~Vaskonen,
%``Isocurvature Constraints on Portal Couplings,''
JCAP \textbf{06} (2016), 022
%doi:10.1088/1475-7516/2016/06/022
[arXiv:1601.07733 [astro-ph.CO]].





\bibitem{Turner:1983he}
M.~S.~Turner,
%``Coherent Scalar Field Oscillations in an Expanding Universe,''
Phys. Rev. D \textbf{28} (1983), 1243.
%doi:10.1103/PhysRevD.28.1243

\bibitem{Martin:2010kz}
J.~Martin and C.~Ringeval,
%``First CMB Constraints on the Inflationary Reheating Temperature,''
Phys. Rev. D \textbf{82} (2010), 023511
%doi:10.1103/PhysRevD.82.023511
[arXiv:1004.5525 [astro-ph.CO]].

\bibitem{Shtanov:1994ce}
Y.~Shtanov, J.~H.~Traschen and R.~H.~Brandenberger,
%``Universe reheating after inflation,''
Phys. Rev. D \textbf{51} (1995), 5438-5455
%doi:10.1103/PhysRevD.51.5438
[arXiv:hep-ph/9407247 [hep-ph]].



\bibitem{Ichikawa:2008ne}
K.~Ichikawa, T.~Suyama, T.~Takahashi and M.~Yamaguchi,
%``Primordial Curvature Fluctuation and Its Non-Gaussianity in Models with Modulated Reheating,''
Phys. Rev. D \textbf{78} (2008), 063545
%doi:10.1103/PhysRevD.78.063545
[arXiv:0807.3988 [astro-ph]].





\bibitem{Kofman:1997yn}
L.~Kofman, A.~D.~Linde and A.~A.~Starobinsky,
%``Towards the theory of reheating after inflation,''
Phys. Rev. D \textbf{56} (1997), 3258-3295
%doi:10.1103/PhysRevD.56.3258
[arXiv:hep-ph/9704452 [hep-ph]].

\bibitem{Greene:1997fu}
P.~B.~Greene, L.~Kofman, A.~D.~Linde and A.~A.~Starobinsky,
%``Structure of resonance in preheating after inflation,''
Phys. Rev. D \textbf{56} (1997), 6175-6192
%doi:10.1103/PhysRevD.56.6175
[arXiv:hep-ph/9705347 [hep-ph]].

\bibitem{Hasegawa:2019jsa}
T.~Hasegawa, N.~Hiroshima, K.~Kohri, R.~S.~L.~Hansen, T.~Tram and S.~Hannestad,
%``MeV-scale reheating temperature and thermalization of oscillating neutrinos by radiative and hadronic decays of massive particles,''
JCAP \textbf{12}, 012 (2019)
%doi:10.1088/1475-7516/2019/12/012
[arXiv:1908.10189 [hep-ph]].

\bibitem{Fields:2019pfx}
B.~D.~Fields, K.~A.~Olive, T.~H.~Yeh and C.~Young,
%``Big-Bang Nucleosynthesis after Planck,''
JCAP \textbf{03}, 010 (2020)
[erratum: JCAP \textbf{11}, E02 (2020)]
%doi:10.1088/1475-7516/2020/03/010
[arXiv:1912.01132 [astro-ph.CO]];
%\cite{Yeh:2020mgl}
%\bibitem{Yeh:2020mgl}
T.~H.~Yeh, K.~A.~Olive and B.~D.~Fields,
%``The Impact of New d(p,\textbackslash{}gamma)He3 Rates on Big Bang Nucleosynthesis,''
[arXiv:2011.13874 [astro-ph.CO]].
%0 citations counted in INSPIRE as of 14 Dec 2020

\bibitem{Elahi:2014fsa}
F.~Elahi, C.~Kolda and J.~Unwin,
%``UltraViolet Freeze-in,''
JHEP \textbf{03}, 048 (2015)
%doi:10.1007/JHEP03(2015)048
[arXiv:1410.6157 [hep-ph]].
%89 citations counted in INSPIRE as of 16 Feb 2021

%\cite{Ellis:2013nka}
\bibitem{eno9}
J.~Ellis, D.~V.~Nanopoulos and K.~A.~Olive,
%``A no-scale supergravity framework for sub-Planckian physics,''
Phys. Rev. D \textbf{89}, no.4, 043502 (2014)
%doi:10.1103/PhysRevD.89.043502
[arXiv:1310.4770 [hep-ph]].
%53 citations counted in INSPIRE as of 11 Dec 2020

%\cite{Greene:1998nh}
\bibitem{Greene:1998nh}
P.~B.~Greene and L.~Kofman,
%``Preheating of fermions,''
Phys. Lett. B \textbf{448}, 6-12 (1999)
%doi:10.1016/S0370-2693(99)00020-9
[arXiv:hep-ph/9807339 [hep-ph]].
%205 citations counted in INSPIRE as of 10 Dec 2020

%\cite{Dufaux:2006ee}
\bibitem{Dufaux:2006ee}
J.~F.~Dufaux, G.~N.~Felder, L.~Kofman, M.~Peloso and D.~Podolsky,
%``Preheating with trilinear interactions: Tachyonic resonance,''
JCAP \textbf{07}, 006 (2006)
%doi:10.1088/1475-7516/2006/07/006
[arXiv:hep-ph/0602144 [hep-ph]].
%134 citations counted in INSPIRE as of 14 Dec 2020

%\cite{Yokoyama:2005dv}
\bibitem{Yokoyama:2005dv}
J.~Yokoyama,
%``Can oscillating scalar fields decay into particles with a large thermal mass?,''
Phys. Lett. B \textbf{635}, 66-71 (2006)
%doi:10.1016/j.physletb.2006.02.039
[arXiv:hep-ph/0510091 [hep-ph]];
%42 citations counted in INSPIRE as of 14 Dec 2020
%\cite{Mukaida:2012bz}
%\bibitem{Mukaida:2012bz}
K.~Mukaida and K.~Nakayama,
%``Dissipative Effects on Reheating after Inflation,''
JCAP \textbf{03}, 002 (2013)
%doi:10.1088/1475-7516/2013/03/002
[arXiv:1212.4985 [hep-ph]].
%51 citations counted in INSPIRE as of 14 Dec 2020


\bibitem{Lozanov:2016hid}
K.~D.~Lozanov and M.~A.~Amin,
%``Equation of State and Duration to Radiation Domination after Inflation,''
Phys. Rev. Lett. \textbf{119}, no.6, 061301 (2017)
%doi:10.1103/PhysRevLett.119.061301
[arXiv:1608.01213 [astro-ph.CO]].

\bibitem{Lozanov:2017hjm}
K.~D.~Lozanov and M.~A.~Amin,
%``Self-resonance after inflation: oscillons, transients and radiation domination,''
Phys. Rev. D \textbf{97}, no.2, 023533 (2018)
%doi:10.1103/PhysRevD.97.023533
[arXiv:1710.06851 [astro-ph.CO]].

\bibitem{Maity:2018qhi}
D.~Maity and P.~Saha,
%``(P)reheating after minimal Plateau Inflation and constraints from CMB,''
JCAP \textbf{07}, 018 (2019)
%doi:10.1088/1475-7516/2019/07/018
[arXiv:1811.11173 [astro-ph.CO]].

\bibitem{Nurmi:2015ema}
S.~Nurmi, T.~Tenkanen and K.~Tuominen,
%``Inflationary Imprints on Dark Matter,''
JCAP \textbf{11}, 001 (2015)
%doi:10.1088/1475-7516/2015/11/001
[arXiv:1506.04048 [astro-ph.CO]].







\bibitem{Ellis:2015pla}
J.~Ellis, M.~A.~G.~Garcia, D.~V.~Nanopoulos and K.~A.~Olive,
%``Calculations of Inflaton Decays and Reheating: with Applications to No-Scale Inflation Models,''
JCAP \textbf{07}, 050 (2015)
%doi:10.1088/1475-7516/2015/07/050
[arXiv:1505.06986 [hep-ph]].




\bibitem{Aghanim:2018eyx}
N.~Aghanim \textit{et al.} [Planck],
%``Planck 2018 results. VI. Cosmological parameters,''
Astron. Astrophys. \textbf{641}, A6 (2020)
%doi:10.1051/0004-6361/201833910
[arXiv:1807.06209 [astro-ph.CO]].

\bibitem{Liddle:2003as}
A.~R.~Liddle and S.~M.~Leach,
%``How long before the end of inflation were observable perturbations produced?,''
Phys. Rev. D \textbf{68}, 103503 (2003)
%doi:10.1103/PhysRevD.68.103503
[arXiv:astro-ph/0305263 [astro-ph]].
%474 citations counted in INSPIRE as of 10 Dec 2020

\bibitem{Fixsen:2009ug}
D.~J.~Fixsen,
%``The Temperature of the Cosmic Microwave Background,''
Astrophys. J. \textbf{707}, 916-920 (2009)
%doi:10.1088/0004-637X/707/2/916
[arXiv:0911.1955 [astro-ph.CO]].
%593 citations counted in INSPIRE as of 10 Dec 2020


\bibitem{Maity:2018exj}
D.~Maity and P.~Saha,
%``CMB constraints on dark matter phenomenology via reheating in Minimal plateau inflation,''
Phys. Dark Univ. \textbf{25}, 100317 (2019)
%doi:10.1016/j.dark.2019.100317
[arXiv:1804.10115 [hep-ph]].


\end{thebibliography}

\end{document}